\documentclass[twocolumn]{aastex62}

\usepackage{amsmath}
\usepackage{graphicx}
\usepackage{amssymb}

\usepackage{color}
\usepackage{subfigure}
\providecommand{\apjs}{Astrophys. J. Suppl.}

\providecommand{\prep}{Phys. Rep.}

\begin{document}

\bibliographystyle{apj}

\title{{\bf Core-collapse Supernova Explosions Driven by the Hadron-quark Phase Transition as a Rare $r$-process Site}}

\author[0000-0003-2479-344X]{Tobias~Fischer$^{*,}$}
\affil{Institute of Theoretical Physics, University of Wroc{\l}aw, 50-204 Wroc{\l}aw, Poland}
\email{$^*$ tobias.fischer@uwr.edu.pl}

\author[0000-0003-4960-8706]{Meng-Ru~Wu$^{\dagger,}$}
\affil{Institute of Physics, Academia Sinica, Taipei, 11529, Taiwan}
\affil{Institute of Astronomy and Astrophysics, Academia Sinica, Taipei, 10617, Taiwan}
\affil{Physics Division, National Center for Theoretical Sciences, Hsinchu, 30013, Taiwan}
\email{$^\dagger$ mwu@gate.sinica.edu.tw}

\author{Benjamin~Wehmeyer}
\affil{Research Centre for Astronomy and Earth Sciences, Konkoly Observatory, 1121 Budapest, Hungary}
\affil{Centre for Astrophysics Research, University of Hertfordshire, Hatfield AL10 9AB, UK}

\author{Niels-Uwe~F.~Bastian}
\affil{Institute of Theoretical Physics, University of Wroc{\l}aw, 50-204 Wroc{\l}aw, Poland}

\author{Gabriel~Mart{\'i}nez-Pinedo}
\affil{GSI Helmholtzzentrum f\"ur Schwerionenforschung, D-64291 Darmstadt, Germany}
\affil{Institut f{\"u}r Kernphysik (Theoriezentrum), Fachbereich Physik, Technische Universit{\"a}t Darmstadt, D-64298 Darmstadt, Germany}

\author[0000-0002-7256-9330]{Friedrich-Karl~Thielemann}
\affil{GSI Helmholtzzentrum f\"ur Schwerionenforschung, D-64291 Darmstadt, Germany}
\affil{University of Basel, 4056 Basel, Switzerland}
\received{{\em 2020 March 2}}\revised{{\em 2020 April 2}}\accepted{{\em 2002 April 2}} \published{{\em 2020 April 30}}

\begin{abstract}
Supernova explosions of massive stars are one of the primary sites for the production of the elements in the universe. Up to now, stars with zero-age main-sequence masses in the range of 35--50~$M_\odot$ had mostly been representing the failed supernova explosion branch. In contrast, it has been demonstrated recently that the appearance of exotic phases of hot and dense matter, associated with a sufficiently strong phase transition from nuclear matter to the quark-gluon plasma at high baryon density, can trigger supernova explosions of such massive supergiant. Here, we present the first results obtained from an extensive nucleosynthesis analysis for material being ejected from the surface of the newly born proto-neutron star of such supernova explosions. 
These ejecta contain an early neutron-rich component and a late-time high-entropy neutrino-driven wind. The nucleosynthesis robustly overcomes the production of nuclei associated with the second $r$-process peak, at nuclear mass number $A\simeq 130$, and proceeds beyond the formation of the third peak ($A\simeq 195$) to the actinides. These yields may account for metal-poor star observations concerning $r$-process elements such as strontium and europium in the Galaxy at low metalicity, while the actinide yields suggests that this source may be a candidate contributing to the abundances of radioactive $^{244}$Pu measured in deep-sea sediments on Earth. \\
\small {\em Unified Astronomy Thesaurus concepts:} {\color{blue}Supernovae(1668); Nucleosynthesis(1131); Nuclear astrophysics(1129);} \\ \\
\end{abstract}


\section{Introduction}
\label{intro}
Core-collapse supernova (SN) explosions have long been considered as the main site for the nucleosynthesis of the elements in the universe. Explosive silicon and/or oxygen burning, associated with the propagation of the SN shock through the stellar envelope, is the main source of intermediate mass and a fraction of iron-group nuclei in the universe, in combination with thermonuclear SN (type Ia). The production of elements heavier than iron has been traditionally associated with the neutrino-driven wind (NDW). This low-mass outflow is ejected via neutrino heating from the surface of the newly formed proto-neutron star (PNS) during the deleptonization phase which starts after the SN explosion is launched, i.e. the explosion shock expands continuously to increasingly larger radii.

In \citet{Qian:1996xt} it has been demonstrated that the conditions of the NDW are entirely determined by (i) properties of the PNS, for example, in terms of radius and mass that depend on the SN evolution prior to the explosion's onset and on the equation of state (EOS), and by (ii) the properties and conditions of neutrino decoupling from matter at the PNS surface. The latter are given by neutrino diffusion from the PNS interior on a timescale of 10~seconds, which leads to the deleptonization of the PNS. Both aspects {\em (i)} and {\em (ii)} are correlated: for example, the weak rates that drive the deleptonization depend on the EOS \citep[see, e.g.,][]{MartinezPinedo:2012,Roberts:2012} while the rate of the deleptonization determines the contraction behavior of the PNS.

Present models of the PNS deleptonization based on three-flavor Boltzmann neutrino transport fail to yield $r$-process conditions, generally having too small neutron excess in combination with generally too low entropy per baryon that prevents any strong $r$-process nucleosynthesis. This led to the consideration that binary neutron star mergers could be the major $r$-process site \citep[for some recent work, see][and references therein]{Goriely:2011,Korobkin:2012uy,Fernandez:2013tya,Wanajo:2014wha,Just:2014fka,Wu:2016pnw,Radice:2016,Siegel:2017nub}. However, there are large uncertainties, in particular, the mass ejection and the timescale until merger are presently still under debate. Relatedly, it is still unclear if binary neutron star mergers can account for the Galactic $r$-process enrichment at low metalicity \citep[cf.][]{Argast:2004,Cowan:2006,Sneden:2008,Wehmeyer:2015}, the europium-to-iron abundance evolution at high metalicity\citep[][]{Komiya:2016xx,Hotokezaka:2018aui,Cote:2018qku}, and if the age--metalicity correlation generally holds \citep[][]{Hirai:2015}. It points to the necessity of an $r$-process site at low metalicity related to core-collapse SNe. Therefore, rare and still speculative sites for the $r$ process associated with massive star explosions have been considered, such as magnetically driven jet-like explosions \citep[][]{Winteler:2012}. However, the strong magnetic fields required for this mechanism to operate and the possible growth of hydrodynamics instabilities, which are likely to dump the formation of jets \citep[for details, cf.][]{Moesta:2014}, are presently under investigation. Moreover, the inclusion of neutrino transport in such simulations may challenge the present results regarding the $r$ process.

The NDW nucleosynthesis depends sensitively on the following three properties, expansion timescale $\tau$, entropy per baryon $S$ and electron fraction $Y_e$. The latter is equivalent to the proton-to-baryon ratio~\citep[for more details, cf.][]{Qian.Wasserburg:2007}. The originally found high-entropy NDW scenario of \citet{Woosley:1994ux}, with $S\simeq100-300~k_{\rm B}$ and subsequently the NDW were considered to be robust $r$-process sites with the production of heavy elements up to the third peak with nuclear mass number $A\sim$195, was soon after ruled out by \citet{Witti:1994} and \citet{Takahashi:1994yz}. Besides the entropy, $Y_e$ is determined by the spectral difference between $\bar\nu_e$ and $\nu_e$, which in turn defines whether the nucleosynthesis path proceeds under neutron-rich ($Y_e<0.5$) or proton-rich ($Y_e>0.5$) conditions. Therefore, neutrino transport is the key input physics in studies that aim to predict the nucleosynthesis outcome.

The numerical studies of \citet{Fischer:2009af} and \citet{Huedepohl:2010} were based on three-flavor Boltzmann neutrino transport, where generally small spectral differences between $\bar\nu_e$ and $\nu_e$ were found to favor proton-rich conditions. With the medium modifications at the level of the mean-field description of \citet{Reddy:1998} for the charged-current weak rates, implemented in \citet{MartinezPinedo:2012} and \citet{Roberts:2012}, \citet{MartinezPinedo:2014} revealed that the nucleosynthesis of elements heavier than molybdenum, with atomic number $Z$=42, cannot be obtained in canonical NDW from PNS connected with neutrino-driven SN explosions \citep[see also][where in addition medium modifications for the $N$--$N$ bremsstrahlung were considered]{Bartl:2016,Fischer:2016b}. These findings also reduce the possibility of nucleosynthesis associated with the $\nu p$ process \citep[for details, cf.][]{Pruet:2005,Pruet:2006,Froehlich:2005ys}, which requires $Y_e>0.5$ in the presence of high $\bar\nu_e$ luminosities. Moreover, the inclusion of weak physics beyond the mean-field level, e.g., nuclear many-body correlations,\footnote{Many-body correlations were taken into account in the study of \citet{Huedepohl:2010} for charged-current absorption and neutral-current scattering processes following the treatment of \citet{Burrows:1998}.} is unlikely to alter these insights. Similar nucleosynthesis results are obtained by \citet{Wanajo:2009}, related to the class of electron-capture SN \citep[][]{Kitaura:2006}, due to the fast ejection of neutron-rich pockets early after the SN explosion's onset, aided by multidimensional SN simulations, which tends to increase the neutron excess \citep[cf.][]{Wanajo:2011a}.

In this article we present the detailed nucleosynthesis analysis of the recent results of massive star explosions associated with supergiant stars of zero-age main-sequence (ZAMS) masses on the order of 35--50~$M_\odot$~\citep[][]{Fischer:2018} as a {\em rare} $r$-process site. Such core-collapse SN can be launched if a first-order phase transition takes place at high baryon density, in excess of normal nuclear density ($\rho_{\rm sat}\simeq2.5\times10^{14}$~g~cm$^{-3}$), in the PNS interior during the post-bounce evolution of such massive stars.  The resulting energetic explosions produce smaller amount of iron-group elements than those obtained in normal core-collapse SN. The associated conditions of mass ejection, which give rise to the nucleosynthesis, will be discussed in the present paper. 

In Section~\ref{sec2} of this article, we review the newly developed class of hadron-quark hybrid EOS together with the associated SN explosions, with a focus on the comparison to the canonical picture of mass ejection from SN simulations based on purely hadronic EOS. Section~\ref{sec3} discusses the conditions and results of the nucleosynthesis analysis. The obtained yields are discussed in comparison with metal-poor star observations as well as those for $^{244}$Pu are the foundation for the discussion presented in Section~\ref{sec:4}. The manuscript closes with a summary in Section~\ref{sec:summary}.

\section{Supernova Explosion Model}
\label{sec2}
The SN simulations discussed in the present section are launched from the stellar progenitor with ZAMS mass of 40~$M_\odot$, from the solar-metallicity stellar evolution series of \citet{Rauscher:2002}; labelled '...a28'. We define this SN model as reference case. Further below, in Sec.~\ref{sec:2.4}, in order to study the sensitivity of the SN explosion on the stellar progenitor, further SN simulation will be discussed that are launched from the 35~$M_\odot$ progenitor of \citet{Rauscher:2002}; belonging to the same series labelled '...a28', the 40~$M_\odot$ progenitor of \citet{Woosley:1995ip}, and the 50~$M_\odot$ progenitor of \citet{Umeda:2007wk}. Except for the 40~$M_\odot$ progenitor of \citet{Woosley:1995ip}, all other progenitor models feature a low post-bounce mass-accretion rate. We will return to this aspect in Sec.~\ref{sec:4.2}. In the following we will discuss and illustrate the SN evolution of the 40~$M_\odot$ reference model.

The SN explosions are triggered by the phase transition from ordinary nuclear matter with hadronic degrees of freedom to the quark-gluon plasma, during the SN post-bounce mass accretion phase. These simulations were performed with the SN model {\tt AGILE-BOLTZTRAN}. It is based on general relativistic neutrino-radiation hydrodynamics in spherical symmetry, featuring three-flavor Boltzmann neutrino transport \citep[cf.][and references therein]{Mezzacappa:1993gm, Liebendoerfer:2004}. Table~1 in \citet{Fischer:2016b} lists the set of weak processes together with the references used in the simulation setup of \citet{Fischer:2018}. Note that we include here the (inverse)neutron decay as additional opacity source for $\bar\nu_e$, which tends to turn material to the proton-rich side during the PNS deleptonization as was discovered recently in \citet{Fischer:2020}.

\subsection{Hadron-quark Hybrid Equation of State}
The new class of hybrid EOS, which was developed for the study of \citet{Fischer:2018}, is based on the common two-phase approach. It employs the relativistic mean field DD2F EOS of \citet{Typel:2009sy} for hadronic matter with density-dependent meson-nucleon couplings, and the {\em string-flip} (SF) microscopic quark-matter model of \cite{Blaschke:2017} which was motivated from Quantum Chromodynamics (QCD). Further details can be found in \citet{Horowitz:1985} and \citet{Blaschke:1986}. Repulsive vector interactions are included \citep[see][and references therein]{Benic:2014jia,Klaehn:2015}. They give rise to additional pressure contributions with increasing density, which is essential in order to be consistent with the existence of massive neutron stars with $\gtrsim$2~$M_\odot$ \cite{Fonseca:2016,Antoniadis:2013,Cromartie:2019}.

In this work we employ the parameterizations DD2F-SF-2, see  Table~I in the supplementary material of \citet{Bauswein:2019}. We note here that this hybrid DD2F-SF EOS was used in the study of binary neutron star mergers of \citet{Bauswein:2019}, which gave rise to a strong signature of a first-order phase transition in the gravitational wave signal. The corresponding cold neutron-star (or hybrid star\footnote{Hybrid stars are neutron (or proto-neutron) stars with quark-matter cores.}) mass-radius relations are given in Fig.~2 of the supplementary material of \citet{Bauswein:2019}, with an onset mass for quark matter at 1.37~$M_\odot$ and a maximum hybrid star mass of 2.16~$M_\odot$. In Table~\ref{tab:eos} we list selected properties of the DD2F-SF-2 EOS at $T=0$ and $\beta$-equilibrium as well as at $T=50$~MeV and for a fixed proton fraction of $Y_p=0.3$.

\begin{table}[t!]
\centering
\caption{Selected properties of the hadron-quark hybrid EOS DD2F-SF-2.}
\begin{tabular}{c ccccccc c cccc c cccc c}
\hline\hline
&&&&&&&& $T$ &&&&& $\rho_{\rm onset}^{\rm a}$ &&&&& $\rho_{\rm final}^{\rm b}$  \\
&&&&&&&& $($MeV$)$ &&&&& $(\rho_{\rm sat})$ &&&&& $(\rho_{\rm sat})$ \\
\hline
$\beta$-equilibrium &&&&&&&& 0 &&&&& 3.1 &&&&& 3.5 \\
$Y_p=0.3$ &&&&&&&& 50 &&&&& 1.0 &&&&& 3.1 \\
\hline
\hline
\end{tabular}
\\
\begin{flushleft}
{\bf Notes.}\\
$^{\rm a}$~Phase transition onset density. \\
$^{\rm b}$~Density corresponding to the onset of pure quark matter.
\end{flushleft}
\label{tab:eos}
\end{table}

The key feature for the SN evolution is the first-order phase transition and the subsequent release of latent heat. For the phase transition from DD2F-to-SF a Maxwell construction is applied. It results in the typical pressure plateau in the hadron-quark mixed phase for symmetric matter, i.e. at proton fraction of $Y_p=0.5$, and at $T=0$. The corresponding phase transition onset(final) densities are listed in Table~\ref{tab:eos}. At finite temperature and arbitrary $Y_p$, the phase transition region widens substantially (see Table~\ref{tab:eos}). Moreover, the phase transition resembles more closely a Gibb's construction due to the presence of more than one conserved charge \citep[details about the construction of phase a transition in the presence of more than one conserved charge can be found in][]{Hempel:2009vp}. It results in a finite pressure slope also in the hadron-quark mixed phase. Recently, it has been found in the studies of \citet{Yasutake:2014} and \citet{Yasutake:2019} that taking into account the more realistically the appearance of different shapes during the phase transition, e.g., bubbles, rods and slabs, collectively known as pasta phases, would resemble more closely such a phase transition construction.

\begin{figure}[t!]
\centering
\includegraphics[width=1.0\columnwidth]{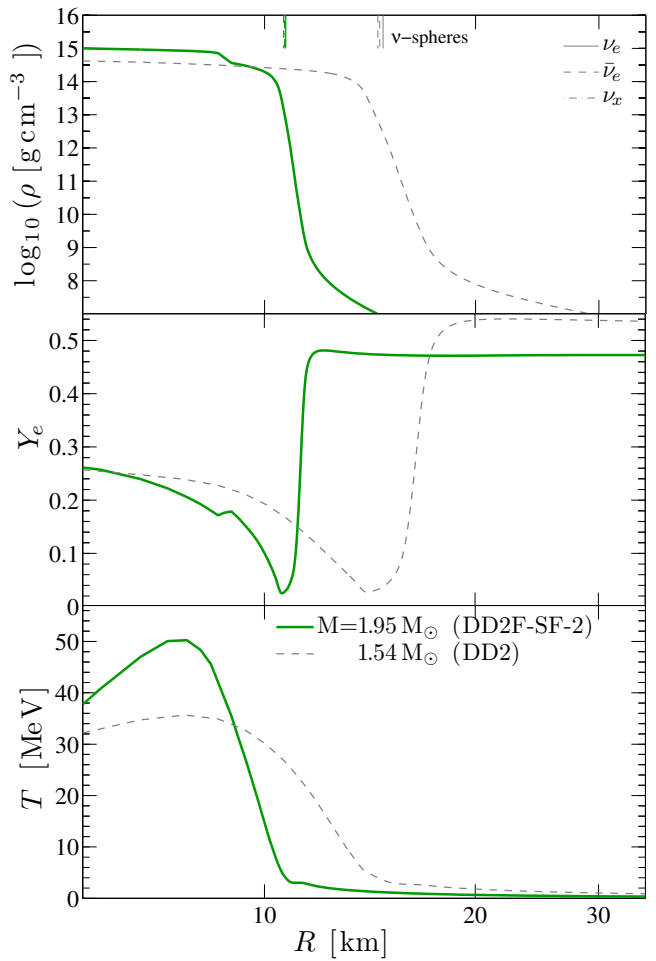}
\caption{PNS structure showing restmass density $\rho$ (top panel), electron abundance $Y_e$ (middle panel) and temperature $T$ profiles (bottom panel) with respect to the radius $R$, at about 5~second after the SN explosion onset was launched with the continuous acceleration of the second shock originating from the hadron-quark phase transition, comparing the 40~$M_\odot$ SN simulation with hadron-quark DD2F-SF-2 hybrid EOS (solid green lines) and a PNS from a neutrino-driven 18~$M_\odot$ SN explosion with the DD2 hadronic EOS (grey dashed lines) from \citet{Fischer:2020}. The thin vertical lines mark the locations of the neutrinospheres of last scattering for $\nu_e$ (solid), $\bar\nu_e$ (dash-dotted) and heavy lepton flavors ($\mu/\tau$-(anti)neutrinos), collectively denoted as $\nu_x=\nu_{\mu/\tau}$ (dashed).}
\label{fig:structure}
\end{figure}

\subsection{Supernova Explosions Triggered by the Hadron-quark Phase Transition}
In the following, we present a brief description of the supernova explosion mechanism due to a hadron-quark phase transition, additional details are provided in \citet{Fischer:2018}. The presence of a mixed phase between the two stable hadronic and quark phases is essential. Once the central PNS enters this hadron-quark mixed phase,  which is hydrodynamically unstable as a result of the Maxwell construction employed here \citep[see the supplementary material in][for example, the velocity of sound drops sharply]{Fischer:2018}, the PNS starts to collapse. This contraction develops into an adiabatic collapse that proceeds supersonically. The collapsing PNS core bounces back in the pure quark-matter phase due to the stiffening induced from the repulsive vector  interactions~\citep[for details, cf.][]{Benic:2014jia,Klaehn:2015}. This evolution is similar to the stellar core bounce when the stellar core collapse halts at the phase transition to homogeneous nuclear matter, however, in the vicinity of $\rho_{\rm sat}$.

While the outer layers of the PNS are still in the mixed phase and hence collapsing, the abrupt halt of the PNS collapse in the pure quark phase results in the formation of an additional strong hydrodynamic shock wave, in addition to the stalled SN accretion shock front standing at a radius of $\sim$50~km. The propagation of this second shock wave is driven from the release of latent heat as hadronic matter keeps on falling onto the second shock being turned into quark matter. Furthermore, the second shock accelerates when reaching the PNS surface. Matter velocities on the order of the speed of light are reached. The propagation of the shock across the neutrinospheres of last scattering releases an additional millisecond neutrino burst~\citep[see Fig.~3 of][]{Fischer:2018}.

Important here is a generally extended post-bounce mass-accretion period, on the order of 1~second, which results in massive PNS of about 2~$M_\odot$ at the onset of the explosion. Hence, the remnants of such type of SN explosion are massive neutron stars with quark-matter core of 2~$M_\odot$ at birth. Both the strong shock produced by the second collapse and the presence of quark matter at the PNS interior have important consequences for the PNS properties as well as for the nucleosynthesis of the ejecta, which will be further discussed in sec.~\ref{sec3}. 

Fig.~\ref{fig:structure} shows the structure of the PNS for the SN simulations with the DD2F-SF-2 hybrid EOS at about 5~seconds after the SN explosion onset, with a PNS containing about 1.95~$M_\odot$ of baryonic mass. In comparison to a {\em canonical} neutrino-driven SN explosion \citep[details can be found in][]{Fischer:2016b,Fischer:2020}, launched from the 18~$M_\odot$ progenitor from \citet{Woosley:2002zz} and with the DD2 purely hadronic EOS\footnote{Note that the two hadronic EOS DD2F and DD2 are identical up to a density of about $2.5\times \rho_{\rm sat}$~\citep[see][]{Typel:2009sy}.}, the resulting PNS contains about 1.54~$M_\odot$ of baryonic mass. The latter PNS is much less compact, with a larger radius on the order of 15--20~km (depending on the stage of deleptonization), in comparison to the radius of about 11--12~km of the massive PNS with a quark-matter core. Significant is also the lower central density and lower temperatures, on the order of $\rho_{\rm central}=4.6\times 10^{14}$g~cm$^{-3}$ and about $T\simeq 30$~MeV, respectively, of the purely hadronic simulation, in comparison to $\rho_{\rm central}=1.0\times 10^{15}$g~cm$^{-3}$ and roughly $T\simeq 40-50$~MeV for the PNS with a quark-matter core (see Fig.~\ref{fig:structure}).

Besides the rapid initial shock expansion, which is not observed in neutrino-driven explosions, the evolution of the central PNS proceeds along similar lines through the deleptonization due to the continuous emission of neutrinos of all flavors. These aspects will be discussed in the following secion.
%

%
\begin{figure}[t!]
\centering
\includegraphics[width=1.0\columnwidth]{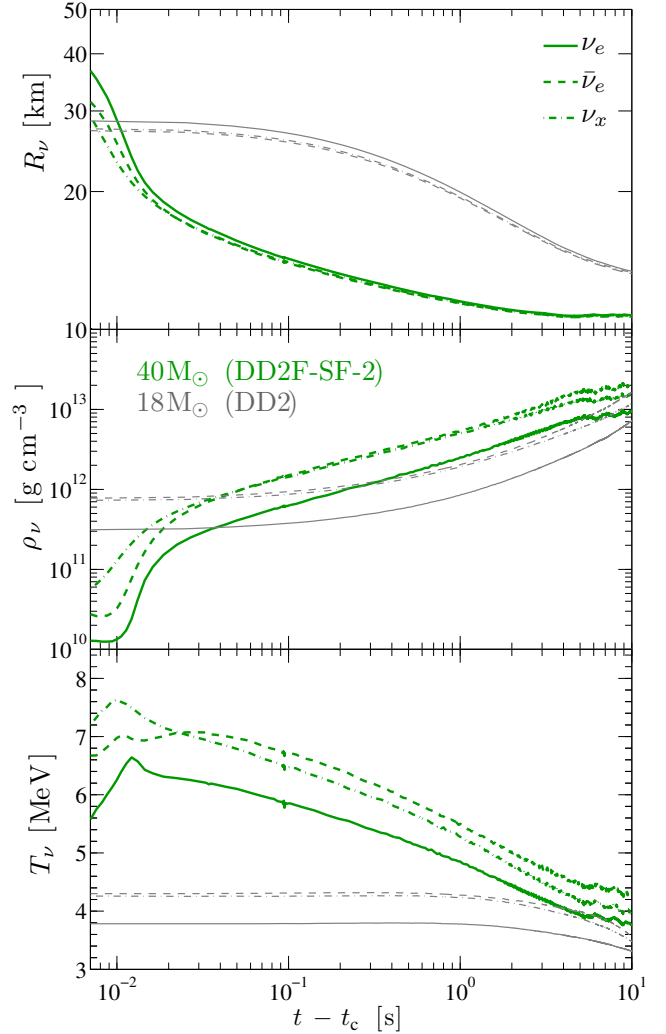}
\caption{Evolution of the neutrinospheres of last scattering, $R_\nu$, in terms of density, $\rho_\nu$, and temperature, $T_\nu$, during the phase of mass ejection, comparing the model associated with the hadron-quark phase transition (thick green lines) of \citet{Fischer:2018} with a massive PNS of 1.95~$M_\odot$ being launched from the 40~$M_\odot$ progenitor of \citet{Rauscher:2002}, and the {\em canonical} neutrino-driven case (thin gray lines) of \citet{Fischer:2020} with an intermediate-mass PNS of 1.54~$M_\odot$ launched from the 18~$M_\odot$ progenitor of \citet{Woosley:2002zz}. We distinguish $\nu_e$ and $\bar\nu_e$, as well as a representative for the heavy lepton flavors $\nu_x=\nu_{\mu/\tau}$. The times are gauged to $t_c$ (see text for definition); for the 18~$M_\odot$ model we select $t_c=0.4$~s corresponding to the moment of the neutrino-driven SN explosion onset when the explosion shock reaches a distance of about $10^3$~km.}
\label{fig:PNS}
\end{figure}

\subsection{Proto-Neutron Star Evolution Beyond the Supernova Explosion Onset}
\label{sec:2.3}
The PNS evolution after the SN explosion has been launched is illustrated in Fig.~\ref{fig:PNS} (green lines), compareing the reference SN simulation launched from the 40~$M_\odot$ solar metallicity progenitor of \citet{Rauscher:2002} (corresponding to the series labelled '...a28') and the PNS evolution of the neutrino-driven explosion of \citet{Fischer:2016b} where the hadronic DD2 model EOS was employed (grey lines in Fig.~\ref{fig:PNS}). Note that all times denoted in the proceeding figures are relative to the SN explosion onset time, $t_c$, which we associate with the PNS collapse due to the hadron-quark phase transition. For the SN simulation based on the hadronic EOS DD2 of \citet{Fischer:2016b}, $t_c$ corresponds to the time when the shock waves crosses a radius of 1000~km. We note that in the quark-matter phase at supersaturation density, the temperatures reach up to 30--60~MeV. Under these conditions, neutrinos are trapped completely. In the Boltzmann neutrino transport equation we employ the hadronic reaction rates. Therefore, we reconstruct the corresponding baryon ($B$) and charge ($C$) chemical potentials from those of up- and down-quarks as follows,
\begin{equation}
\mu_B = \mu_u + 2\,\mu_d\equiv\mu_n\;\;,\;\;\mu_C = \mu_u-\mu_d\equiv\mu_n-\mu_p\;.
\end{equation}
This approximation should hold for times less than 10~s, as long as neutrinos of all energies between 0.5--300~MeV, decouple in the hadronic phases. The associated partial neutron and proton densities are then obtained by the derivative of the thermodynamic pressure, $n_i=\partial P/\partial \mu_i$. Note further that all simulations are based on identical neutrino input physics, i.e. three-flavor Boltzmann neutrino transport with the same set of weak rates \citep[see Table~(1) in][]{Fischer:2016b}, where for the neutrino-nucleon scattering rates inelastic effects and contributions from weak magnetism are taken approximately into account following \citet{Horowitz:2001xf}. The charged-current weak rates are based on the fully inelastic treatment of \citet{Fischer:2020}. Vertex corrections to the $N$--$N$ bremsstrahlung, which tend to suppress the neutrino pair processes towards higher density, are taken into account following the approximate treatment of \citet{Fischer:2016b}.

\begin{figure*}[t!]
\centering
\subfigure[~Hadron-quark phase transition]{\includegraphics[width=0.475\textwidth]{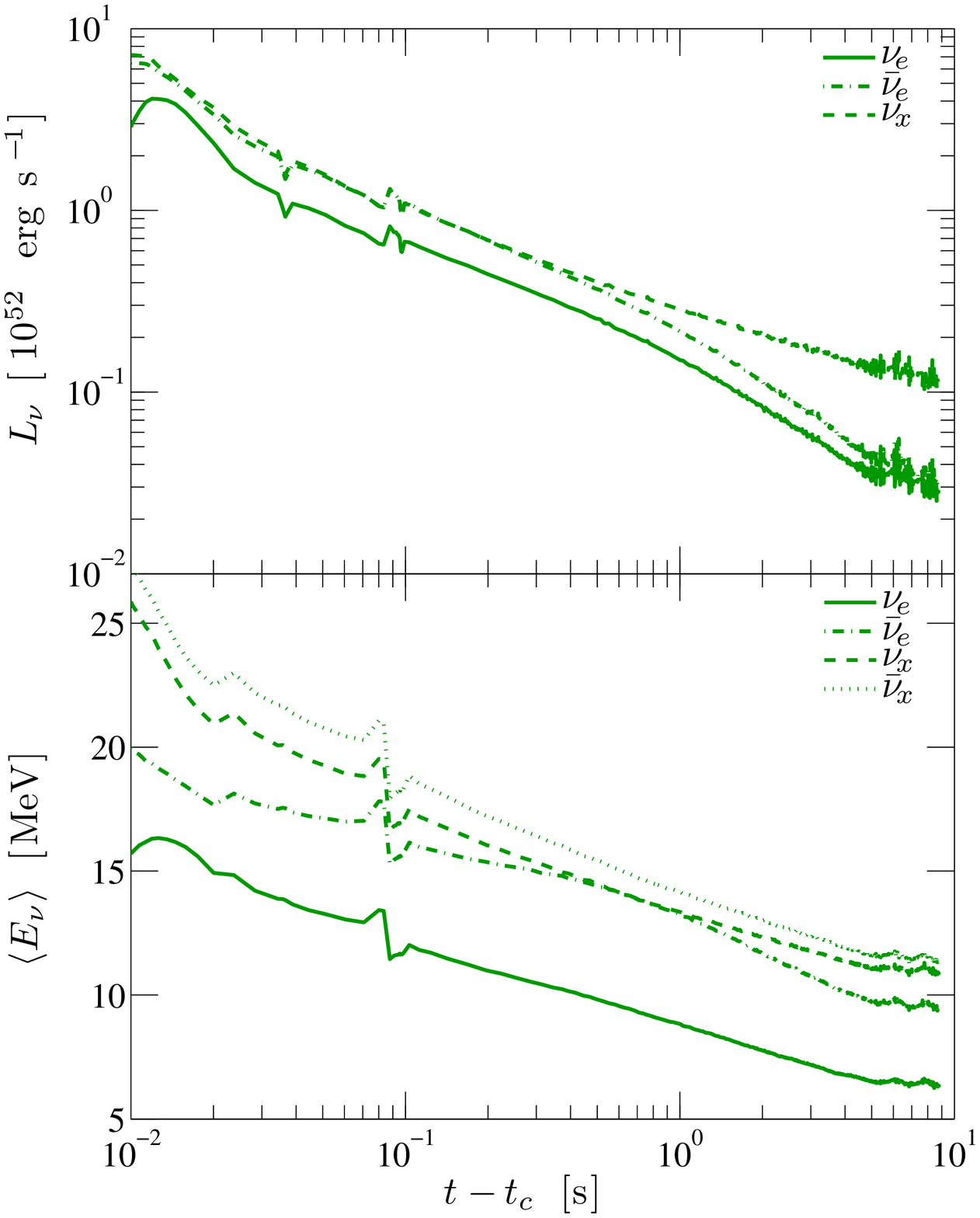}\label{fig:nuLEa}}
\subfigure[~Purely hadronic EOS~\citep{Fischer:2020}]{\includegraphics[width=0.475\textwidth]{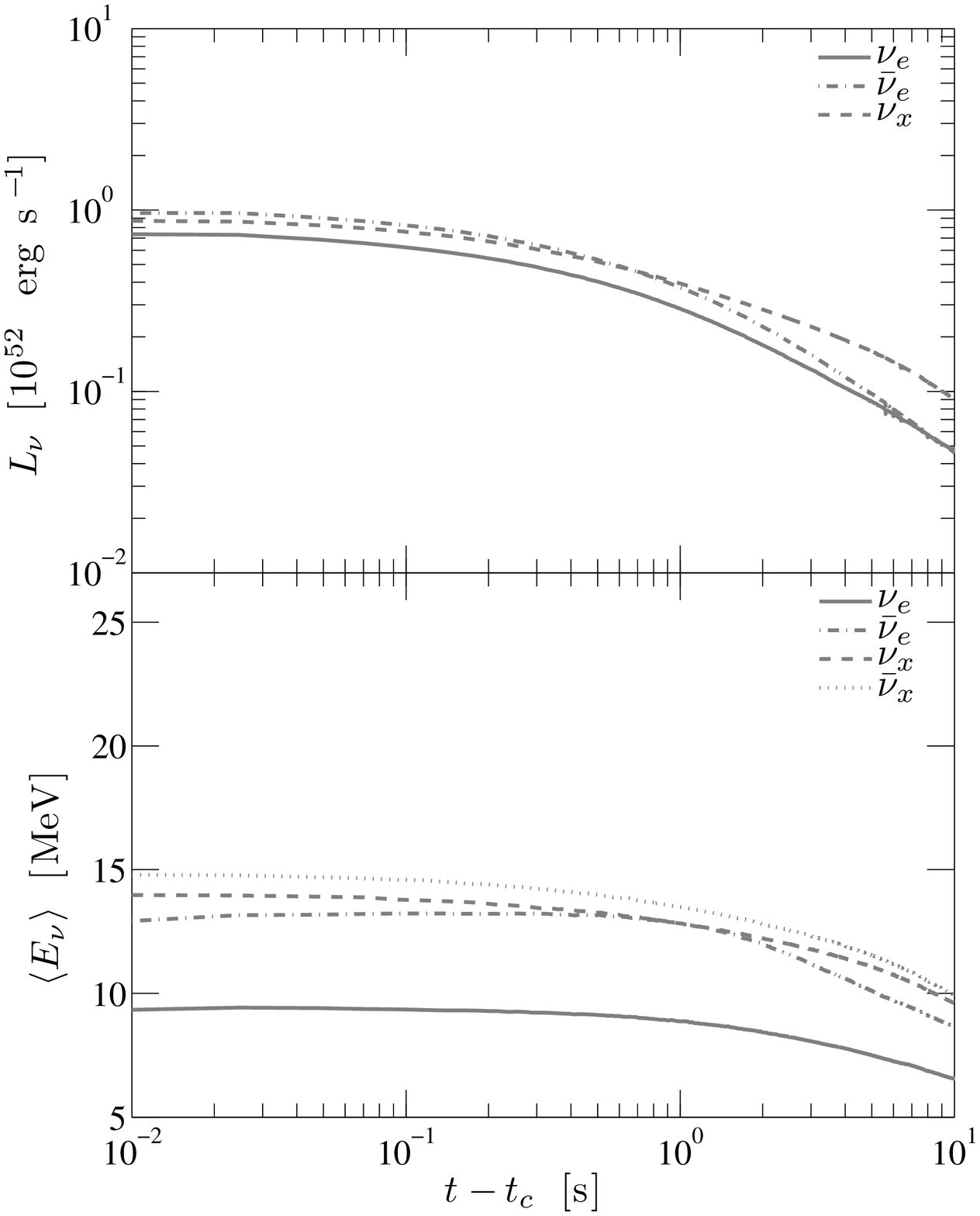}\label{fig:nuLEb}}
\caption{Evolution of the neutrino luminosity $L_\nu$ and average energy $\langle E_\nu \rangle$ for all flavors, sampled in the comoving frame of reference at a distance of 500~km, comparing the SN simulation driven by the hadron-quark phase transition of \citet{Fischer:2018} in graph~(a) with massive PNS of 1.95~$M_\odot$, launched from the 40~$M_\odot$ progenitor, and the canonical neutrino-driven case  of \citet{Fischer:2020} with an intermediate-mass PNS of 1.54~$M_\odot$, launched from the 18~$M_\odot$ in graph~(b). Heavy lepton flavors are collectively denoted as $(\nu_x, \bar\nu_x)$ and as a representative for the heavy lepton flavors we only show the luminosity for $\nu_x$ since they are nearly indistinguishable from $\bar\nu_x$.}
\label{fig:nuLE}
\end{figure*}

The largest difference, comparing the simulations with the hadron-quark DD2F-SF-2 hybrid EOS and the DD2 purely hadronic EOS, arises in the much faster PNS contraction illustrated via the radii $R_{\rm PNS}$. The PNS with quark-matter core is significantly more compact (see also Fig.~\ref{fig:structure}), which in turn leads to PNS radii on the order of 10--12~km already during the early evolution on the order of 1~second after $t_c$. This EOS effect has important consequences for the conditions at the PNS surface, which are particularly relevant at the neutrinospheres shown in Fig.~\ref{fig:PNS}; for example, the significantly higher temperature of the PMS with the quark-matter core, initially $T_\nu\simeq 5-6$~MeV, in comparison to that of the PNS with a purely hadronic EOS with $T_\nu\simeq3.5-4.5$~MeV. Consequently, also, the neutrinospheres are located at generally higher densities (see $\rho_{\nu}$ in Fig.~\ref{fig:PNS}).

The evolution of the neutrino luminosities and average energies is illustrated in Fig.~\ref{fig:nuLEa}, again in comparison to those of the neutrino-driven model from \citet{Fischer:2016b} with the DD2 purely hadronic EOS in Fig.~\ref{fig:nuLEb}. Despite the different behavior at early times, which is due to the 2$^{\rm nd}$ shock passage across the neutrinospheres, the magnitudes of the neutrino luminosities are on the same order. However, the average energies are significantly higher and approach a similar magnitude only towards late times $\sim$10~s. In particular the spectral differences between $\bar\nu_e$ and $\nu_e$ are larger, due to the higher temperatures of the protoneutron star for the hadron-quark case. For the luminosities there is a compensation between the larger energy fluxes found in the hadron-quark case and the smaller emitting area due the more compact PNS. The different PNS interior structure, in particular the higher density and temperature, features a different $Y_e$ profile (see Fig.~\ref{fig:structure}). 

\subsection{Dependence on the Stellar Progenitor}
\label{sec:2.4}
In order to study the sensitivity of the SN explosion scenario associated with the hadron-quark phase transition, in addition to that of the 40~$M_\odot$ progenitor of \citet{Rauscher:2002} discussed above, further SN simulation launched from the 35~$M_\odot$ progenitor of \citet{Rauscher:2002}; belonging to the same series labelled '...a28', the 40~$M_\odot$ progenitor of \citet{Woosley:1995ip}, and the 50~$M_\odot$ progenitor of \citet{Umeda:2007wk}, were performed. Selected quantities of all these simulations are summarized in Table~\ref{tab:prog}. All stellar models are at solar metallicity.

\begin{table}[t!]
\centering
\caption{Summary of the SN simulation results.}
\begin{tabular}{ccccccc}
\hline\hline
$M_{\rm prog}^{\rm a}$ && $\dot{M}_{r=500~\rm km}^{\rm c}$ & $t_c^{\rm b}$  & $t_{\rm explosion}^{\rm d}$ & $M_{\rm NS}^{\rm e}$ & $E_{\rm expl}^{\rm f}$ \\
$(M_{\odot})$ && $(M_{\odot}~\rm s^{-1})$ & $($s$)$ &  $($s$)$ & $(M_{\odot})$ & $(10^{51}$~erg$)$ \\
\hline
$35^\dagger$ && 0.32 &1.061  & 1.069 & 1.91 & 1.7 \\
$40^\dagger$ && 0.45 & 0.895  & 0.903 & 1.95 & 2.0 \\
$50^\ddagger$ && 0.53 & 0.825  & 0.832 & 2.02 & 3.8 \\
\hline
$40^*$ && 1.32 & 0.401 & 0.402$^{\rm k}$ & 2.20$^{\rm l}$ & ... \\
\hline
\hline
\end{tabular}
\\
\begin{flushleft}
{\bf Notes.}\\
$^{\rm a}$~ZAMS masses of the stellar series from
$\dagger$:~\citet{Rauscher:2002}, 
$\ddagger$:~\citet{Umeda:2007wk}, $^*$:~\citet{Woosley:1995ip}. \\
$^{\rm b}$~PNS collapse onset time post-bounce. \\
$^{\rm c}$~Mass-accretion rate at a post-bounce time of 0.3~s. \\
$^{\rm d}$~Explosion onset post-bounce time, when $R_{\rm shock}\simeq 10^3$~km. \\
$^{\rm e}$~Remnant neutron star baryon mass obtained after $\sim$10~s. \\
$^{\rm f}$~Diagnostic explosion energy evaluation at  10~s after $t_c$. \\
$^{\rm k}$~Post-bounce time of black hole formation. \\
$^{\rm l}$~Encloses mass at the moment of black hole formation. \\
\end{flushleft}
\label{tab:prog}
\end{table}
\begin{figure*}[t!]
\centering
\includegraphics[width=2\columnwidth]{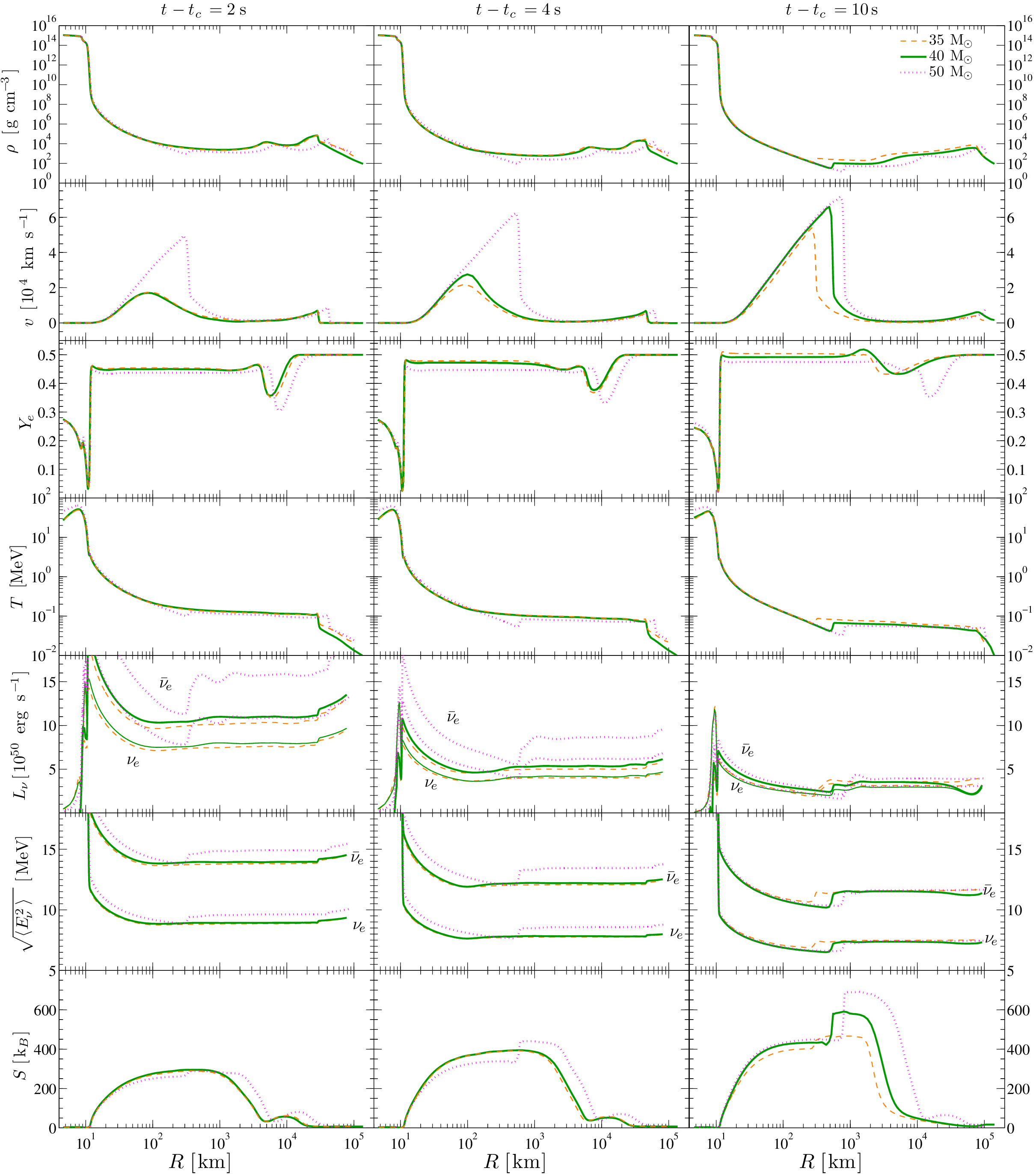}
\caption{Radial profiles of the rest-mass density $\rho$, velocity $v$ with pronounced reverse shocks present, electron fraction $Y_e$, temperature $T$, neutrino luminosity $L_\nu$ for $\nu_e$ and $\bar\nu_e$, their root-mean-square energy $\sqrt{\langle E_\nu^2\rangle}$ and the entropy per particle $S$, comparing the three different initial progenitors under investigation: 35~$M_\odot$ and 40~$M_\odot$ of \citet{Rauscher:2002} as well as the 50~$M_\odot$ of \citet{Umeda:2007wk}. The data correspond to the evolving SN evolution at about 2~s (left panel), 4~s (middle panel), and 10~s (right panel) after the explosion onset.}
\label{fig:fullst}
\end{figure*}

The largest impact of relevance for this study arises due to the different post-bounce mass-accretion rates. Low mass accretion rates, on the order of $\dot M\simeq 0.3-0.5$~$M_\odot$~s$^{-1}$, are found for the progenitors of \citet{Rauscher:2002} and \citet{Umeda:2007wk}. These simulations feature an extended post-bounce mass-accretion period on the order of 0.8--1.0~s, until the conditions for the hadron-quark phase transition are reached at the PNS interior. There is the direct correlation between high (low) $\dot M$ and short (long) $t_{\rm c}$, featuring not only more (less) massive PNS as remnants (see Table~\ref{tab:prog}) but also a direct impact on the dynamics of the ongoing SN explosion, as illustrated in Fig.~\ref{fig:fullst}. The simulation with the longest post-bounce time until the hadron-quark phase transition, launched from the 50~$M_\odot$ progenitor of \citet{Umeda:2007wk}, has the most dynamic NDW phase with highest PNS temperatures, highest reverse-shock velocities and postshock entropy per particle (see Fig.~\ref{fig:fullst}). The two other simulations launched from the 35~$M_\odot$ and 40$_\odot$ progenitor from \citet{Rauscher:2002}, feature longer post-bounce mass accretion periods until the hadron-quark phase transition and hence a less dynamic neutrino-driven wind phase with a less pronounced reverse shock, lower temperatures at the PNS interior, and a lower entropy per particle of the postshocked region. Note the simulation with an extremely high post-bounce mass accretion rate, the 40~$M_\odot$ progenitor from \citet{Woosley:1995ip} with $\dot M>1.0$~$M_\odot$~s$^{-1}$, which exceeds the maximum mass of the DD2F-SF-2 EOS at the moment of the hadron-quark phase transition. As a consequence, the second shock wave formed also in this case cannot expand and trigger the SN explosion onset. Instead, the PNS behind the second shock wave collapses and a black hole forms. The enclosed baryon  mass at this moment is given in Table~\ref{tab:prog}. 

The SN simulation results of the remaining 3 progenitors, 35~$M_\odot$ and 40~$M_\odot$ of \citet{Rauscher:2002} as well as the 50~$M_\odot$ of \citet{Umeda:2007wk} with successful shock revival and subsequent explosions, are illustrated in Fig.~\ref{fig:fullst} at three different times, about 2~seconds (left panel), 4~seconds (middle panel), and 10~seconds (right panel) after the explosion onset. From Fig.~\ref{fig:fullst} one can identify that an earlier(later) phase transition leads to a more(less) energetic explosion. The diagnostic explosion energies listed in Table~\ref{tab:prog} \citep[for the definition, see][and references therein]{Fischer:2009af}, contain  contributions from the gravitational binding energy of the stellar envelopes. Note that they are estimated at about 10~s after the explosion onsets and may differ from the actual explosion energies which are associated with the kinetic energies of the ejecta at the moment of shock break out \citep[cf.][]{Fischer:2018}. Furthermore, the SN simulations with higher(lower) explosion energies feature a lower(higher) $Y_e$ of the NDW, and in turn a slower(longer) deleptonization phase as a consequence of the higher(lower) temperature at the PNS interior. Despite significant differences of the postshock entropy between the different stellar models as well as the timescale for the appearance of the reverse shock \citep[cf.][where the role of the reverse shock on the nucleosynthesis conditions has been studied in]{Fischer:2009af,Arcones:2006uq}, the pre-shock entropy is most relevant for the primary nucleosynthesis path, while the reverse shock may have an effect on modifying the detailed abundance pattern \citep[][]{Arcones:2011c}. In Fig.~\ref{fig:fullst} (bottom panels) we can identify that the pre-shock entropy for all models have similar magnitude with a slightly lower value for the 50~$M_\odot$ model. Hence, it can be expected that most crucial for the nucleosynthesis outcome will be the values obtained for $Y_e$, where we find larger differences indeed between the simulations of the different stellar models. Lowest values of $Y_e$ are obtained in the NDW for the simulations launched from the 50~$M_\odot$ progenitor, due to the largest luminosity and spectral differences obtained between $\bar\nu_e$ and $\nu_e$ (see Fig.~\ref{fig:fullst}).

These differences of the SN explosion dynamics and of the neutrino fluxes and spectra have important consequences on the nucleosynthesis of heavy elements, in particular variations of the electron fraction and entropy per particle, which will be further discussed in the next section. 

\section{Nucleosynthesis of Heavy Elements}
\label{sec3}
In the following we first focus on the SN explosion launched from the 40~$M_\odot$ progenitor of \citet{Rauscher:2002} as the reference case to discuss and illustrate the evolution of the nucleosynthesis relevant conditions as well as the yields from extensive nucleosynthesis calculations, in Sections~\ref{sec:3.1} and Sec.~\ref{sec:3.2}, respectively. Further below, a comparison of results obtained for different progenitor models is given in Sec.~\ref{sec:3.3}.

\begin{figure*}[t!]
\centering
\includegraphics[width=2.\columnwidth]{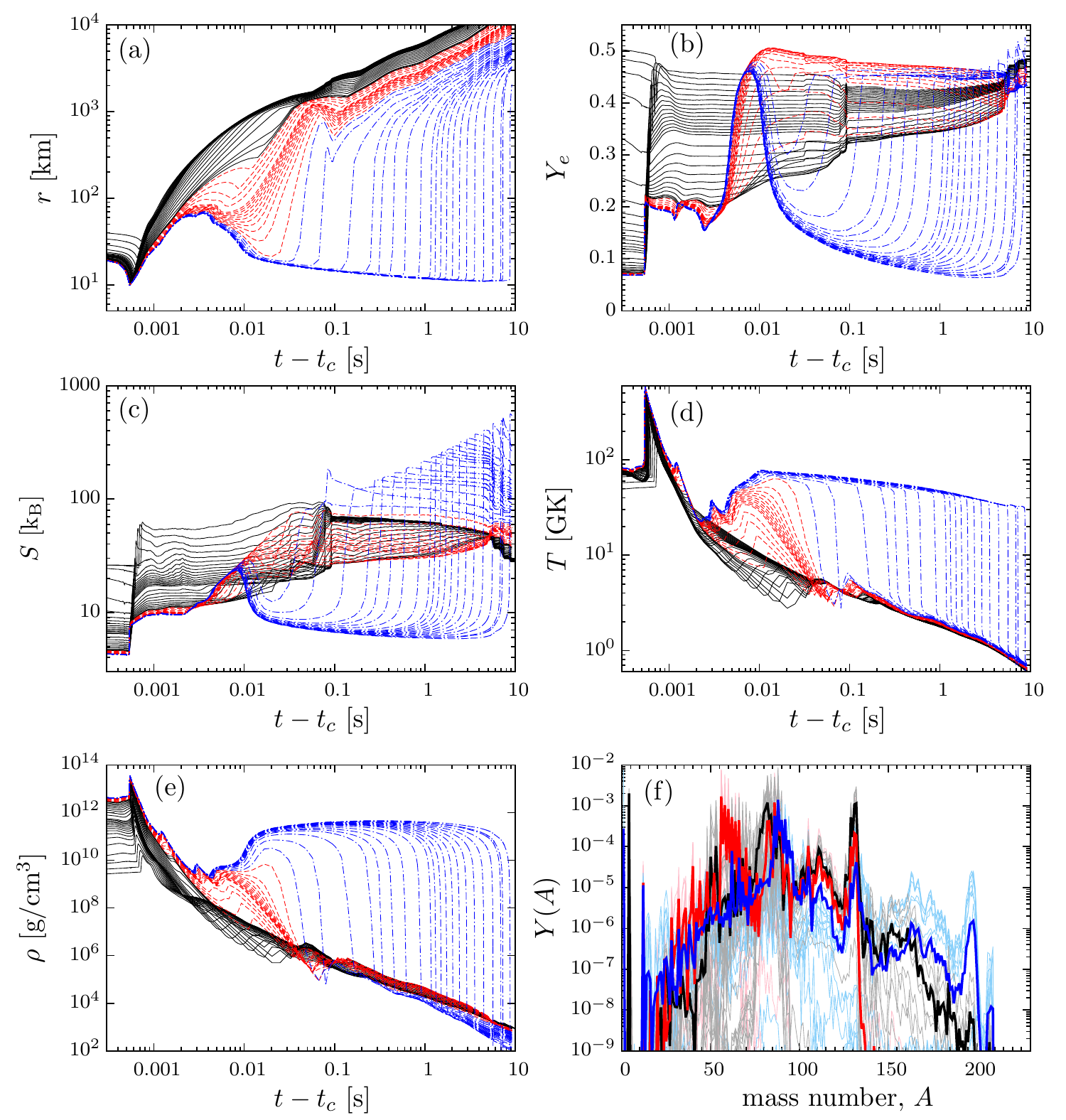}
\caption{Evolution of trajectories representing different ejecta including the direct ejecta (black curves), the intermediate ejecta (red curves), and the neutrino-driven wind (blue curves), showing the radius $r$ in graph~(a), electron fraction $Y_e$ in graph~(b), entropy per nucleon $S$ in graph~(c), temperature $T$ in graph~(d) and the density $\rho$ in graph~(e). Graph~(f) shows the resulting nucleosynthesis abundances $Y(A)$ at the time of 1~Gyr as a function of nuclear mass number $A$ for these trajectories (thin lines), together with $Y(A)$ averaged over all trajectories for the three different kinds of ejecta (thick lines).}
\label{fig:traj}
\end{figure*}

For the nucleosynthesis analysis, $\sim$200 mass trajectories are being selected in a postprocessing fashion from the comoving baryonic mass mesh of {\tt AGILE-BOLTZTRAN}, sampling ejecta driven by the shock from the outer stellar envelope for which the maximum density during the evolution never exceeds $10^{10}$~g~cm$^{-3}$, as well as those inner ones ejected from the PNS surface with density $\gtrsim 10^{10}$~g~cm$^{-3}$ before the shock passage.  We note here that the inner ejecta can be further classified into three subcategories: direct, intermediate, and the NDW, depending on how they become unbound (see Section~\ref{sec:3.1}).

We adopt the nuclear reaction network used in \citet{Wu:2018mvg} to perform nucleosynthesis calculations for all the trajectories. The network uses experimentally measured nuclear masses, decay rates, and reaction rates, supplemented by theoretical predictions for quantities that are not yet determined by experiments. The theoretical masses and the beta decay rates are based on the FRDM mass model \citep{Moller:1993ed} and prediction from \citet{Moller:2003fn}, respectively. Details of other theoretical reactions rates can be found in \citet{Mendoza-Temis:2014mja}. 
For the inner ejecta, the nucleosynthesis computation for each trajectory is performed starting at a temperature of $T=15$~GK with initial abundances well within the nuclear statistical equilibrium (NSE), determined by the following independent thermodynamic conditions $Y_e$, $T$, and $\rho$.  From $T=15$~GK to $T=10$~GK, the composition is evolved assuming NSE but we account for changes in $Y_e$ by $e^{\pm}$ captures and $\nu_e$ and $\bar\nu_e$ absorption on free nucleons. For the neutrino absorption rates, we include the weak magnetism correction factors given by \citet{Horowitz:2001xf} suitable for the conditions in the NDW. For $T<10$~GK, the evolution of the nuclear abundances are then computed with the full network. The temperature and the density evolution of the ejecta are given by the trajectory data from the SN simulation. Beyond the SN simulation time we extrapolate the evolution of density and temperature as $\rho(t) \propto t^{-3}$ and $T(t)\propto t^{-1}$, featuring homologous and adiabatic expansion.

For matter being ejected from the outer stellar envelope that can undergo explosive oxygen and silicon burning, producing the iron-group nuclei, we separate their treatments depending on the maximum postshock temperature $T_{\rm max}$. For those with $T_{\rm max}\geq 10$~GK, the nucleosynthesis computations are performed in a similar way as the inner ejecta by starting at the time when $T(t)={\rm min}(T_{\rm max},15~\rm{GK})$. As for the trajectories with $T_{\rm max}< 10$~GK, we take the detailed isotopic abundances from the progenitor model and start their composition evolution from the beginning of our SN simulation time. Below, we focus on the inner ejecta component consisting of $\sim 10^{-2}$~$M_\odot$ that allows for $r$-process nucleosynthesis to occur.

\subsection{Classification of the Inner Ejecta}\label{sec:3.1}
Differently from the {\em canonical} SN explosions, where the entire inner ejecta are determined by the NDW due to neutrino heating radiated from the PNS, here the passage of the strong second shock due to the hadron-quark phase transition can expel part of the material which belongs to the PNS before the NDW operates. We classify the inner ejecta in the 40~$M_\odot$ reference model into three categories,
\begin{equation*}
\begin{array}{rl}
\rm (I)\,\rm direct~\rm ejecta & \simeq 0.97\times 10^{-2}~M_\odot, \\
\rm (II)\,\rm intermediate~\rm ejecta & \simeq 3.19 \times 10^{-3}~M_\odot~, \\
\rm (III)\,\rm NDW & \simeq 0.43 \times 10^{-3}~ M_\odot~,
\end{array}
\end{equation*}
according to the evolution of the hydrodynamic quantities of the mass elements \citep[see also][]{Nishimura:2012}.  Figures~\ref{fig:traj}~(a)--(e) show the evolution of the tracer mass elements of the inner ejecta, in terms of radius $r$, $Y_e$, entropy per nucleon $S$, temperature $T$, and density $\rho$, after the PNS collapse onset time denoted as $t_c$; These trajectories are marked with different colors. The direct, intermediate, and NDW ejecta are marked by black, red, and blue curves, respectively.

In Figure~\ref{fig:traj}, the panels (a) and (e) clearly display the differences of the radial and density evolution of these three kinds of ejecta. The direct ejecta are driven almost entirely by the second bounce shock and expand continuously outwards. For the intermediate ejecta and the NDW, both have a similar evolution as the direct ejecta during the first $\sim$3~ms after $t_c$, as they are initially part of the PNS surface, which experiences large-scale oscillations due to the central quark-matter phase settling into a quasi-stationary state after the highly dynamical hadron-quark phase transition. However, they fail to be directly ejected by the 2$^{\rm nd}$ shock passage and fall back into the steepening gravitational potential of the oscillating central PNS. The continuous energy deposition of neutrinos emitted from the PNS is able to unbind the intermediate ejecta before them falling back entirely onto the PNS surface ($\rho\simeq 10^{10}$~g~cm$^{-3}$). In contrast, the NDW settle back entirely to the PNS surface and only becomes unbound later exclusively via neutrino heating.

Due to these nonstandard ejection mechanisms, the three ejecta components exhibit distinct thermodynamical histories. For the direct ejecta, the entropy experience a sudden increase to $10$--$60$~$k_{\rm B}$ per baryon (see graph~(c) in Fig.~\ref{fig:traj}) due to the shock heating and the resulting fast expansion.  Also, due to the high postshock temperature, on the order of $T\gtrsim 10$~MeV (graph~(d) in Fig.~\ref{fig:traj}), the $Y_e$ is reset by electron and positron captures to values between $Y_e\simeq0.2-0.46$ (see graph~(b) in Fig.~\ref{fig:traj}).

On the other hand, the entropy and $Y_e$ in both the intermediate ejecta and the NDW are being kept at $S\simeq 10$~$k_{\rm B}$ and $Y_e\simeq0.2$ after the shock passage. For the intermediate ejecta, the subsequent neutrino heating further increases $Y_e$ and $S$. The ones ejected earlier (later) receive smaller (larger) amount of neutrino heating during their expansion and thus have lower (higher) values of $Y_e$ and $S$. Consequently, the intermediate ejecta show large dispersion of $Y_e$ and $S$ during the expansion, ranging between $Y_e\simeq 0.26-0.5$, and $S\simeq 20-100$~$k_{\rm B}$, respectively. For the NDW, since all tracer mass elements fall back to the PNS surface before being ejected by neutrino energy deposition, their $Y_e$ and $S$ evolution are reset and determined by the neutrino luminosities and mean energies~\citep{Qian:1996xt} like the typical NDW. However, as will be discussed in the Sec.~\ref{sec:3.2}, their asymptotic values relevant for the nucleosynthesis differ drastically from the NDW obtained in the typical neutrino-driven SN explosions.

\begin{table*}[t!]
\centering
\caption{Ejecta Properties of the Supernova Explosion Models.}
\begin{tabular}{cccccccccc}
\hline\hline
$M_{\rm prog}^{\rm a}$ & $M_{\rm inner~ejecta}^{\rm b}$ & $M_{\rm direct}^{\rm c}$ & $M_{\rm intermediate}^{\rm d}$ & $M_{\rm NDW}^{\rm e}$ & $M_{\rm Fe}^*$ & $M_{\rm Sr}^*$ & $M_{\rm Eu}^*$ & $M_{^{244}\text{Pu}}^{*,\times}$ & $M_{^{60}\text{Fe}}^{**}$\\
$(M_{\odot})$ & $(10^{-2}~M_{\odot})$ & $(10^{-2}~M_{\odot})$ & $(10^{-3}~M_{\odot})$ &  $(10^{-3}~M_{\odot})$ &
$(10^{-2}~M_{\odot})$ & $(10^{-4}~M_{\odot})$ & $(10^{-6}~M_{\odot})$ &  $(10^{-10}~M_{\odot})$ &  $(10^{-5}~M_{\odot})$ \\
\hline
35$^\dagger$ & 1.44 & 1.03 & 3.72 & 0.33 & 7.15 & 6.50 & 7.86 & 2.09 & 12.2 \\
40$^\dagger$ & 1.33 & 0.97 & 3.19 & 0.43 & 7.62 & 6.15 & 2.26 & 2.64 & 3.2 \\
50$^\ddagger$ & 1.80 & 1.45 & 3.06 & 0.46 & 3.73 & 8.22 & 11.03 & 22.59 & 0.5 \\
\hline
\hline
\end{tabular}
\\
\begin{flushleft}
{\bf Notes.} All yields are evaluated after about 1~Gyrs.\\
$\;^{\rm a}$~ZAMS masses of the stellar models (same as in Table~\ref{tab:prog}), $\dagger$:~\citet{Rauscher:2002} and $\ddagger$:~\citet{Umeda:2007wk}. \\
$\;^{\rm b}$~Total {\em inner ejecta}  mass, launched from the PNS surface. \\
$\;^{\rm c}$~Total {\em direct} ejecta mass. \\
$\;^{\rm d}$~Total {\em intermediate} ejecta mass. \\
$\;^{\rm e}$~Total {\em neutrino-driven wind} ejecta mass. \\
$\;^*$~Ejecta masses for Fe, Sr, Eu, and $^{244}$Pu.\\
$\;^\times$~The amount of $^{244}$Pu is computed at  $\tau_{1/2}(^{248}\rm{Cm})\ll t \ll \tau_{1/2}(^{244}\rm{Pu})$, where $\tau_{1/2}$ labels the half-life of an isotope. \\
$^{**}$~$^{60}$Fe ejecta mass, correspondig to the stellar progenitor model~\citep[][]{Rauscher:2002, Umeda:2007wk}.
\end{flushleft}
\label{tab:ejecta}
\end{table*}
\begin{figure}[t!]
\includegraphics[width=\columnwidth]{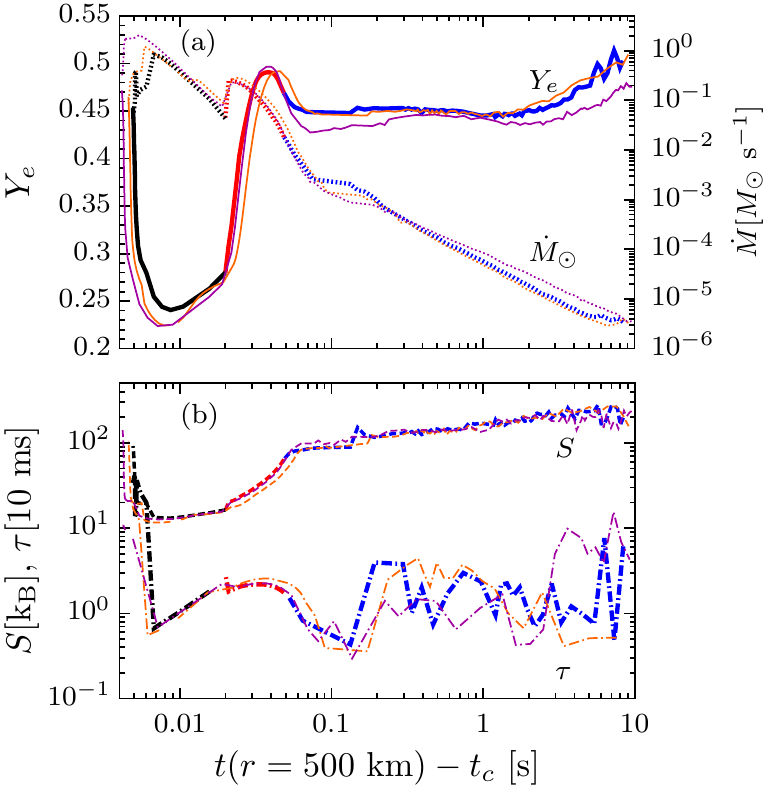}
\caption{Evolution of the nucleosynthesis relevant conditions. Top panel:~$Y_e$ (solid line) and mass-ejection rate $\dot{M}$ (dotted line). Bottom panel: expansion timescale $\tau$ (dashed-dotted line) and entropy per baryon $S$ (dashed line). All quantities are with respect to time after $t_c$ corresponding to the PNS collapse triggered due to the hadron-quark phase transition.  Segments with different colors correspond to different ejecta components for the reference model of $40$~$M_\odot$ (thick lines): the direct ejecta (black), the intermediate ejecta (red), and the neutrino-driven wind (blue). Thin lines show the quantities for the other two SN explosion models launched from the 35~$M_\odot$ (orange color) and the 50~$M_\odot$ (magenta color).}
\label{fig:nucleo_conditions}
\end{figure}

\subsection{Nucleosynthesis Conditions and Yields}\label{sec:3.2}
To connect the differences in the ejection mechanism of the three kinds of ejecta to their nucleosynthesis outcome, we further extract the $S$, $Y_e$, and the expansion timescale $\tau\equiv r/v$ of each tracer element at the moment when the temperature of the ejecta decrease below $T\simeq5$~GK. We show in Fig.~\ref{fig:nucleo_conditions} these quantities, relevant for determining the nucleosynthesis~\citep[cf.][]{Hoffman:1996aj}, together with the mass outflow rate $\dot M=4\pi r^2 \rho v$, as a function of the time when the tracer elements reach $r=500$~km. The corresponding nucleosynthesis abundances $Y(A)$ as a function of mass number $A$ for all trajectories, as well as the values averaged over trajectories inside each kind of ejecta, are shown in Fig.~\ref{fig:traj}(f) by thin and thick curves, respectively.

Fig.~\ref{fig:nucleo_conditions} clearly illustrates the differences for the three kinds of ejecta. The direct ejecta (black segments) feature in general high $\dot M\simeq0.1-1$~$M_\odot$~s$^{-1}$, and contain an early component with rapidly decreasing $Y_e$, from $Y_e\simeq 0.45$ to $Y_e\simeq 0.3$, and decreasing entropy per baryon, from $S\simeq 100~k_{\rm B}$ to $S\simeq 20~k_{\rm B}$, as well as expansion timescale $\tau$ decreasing from $\tau\simeq 100$~ms to $\tau\simeq 10$~ms, within the first $\simeq$5~ms after the PNS collapse time $t_c$ (see Fig.~\ref{fig:nucleo_conditions}). This early component corresponds to matter being ejected from the outermost parts of the PNS, with generally lower baryon density $\rho\simeq 10^{10}$~g~cm$^{-3}$ (see Fig.~\ref{fig:traj}). After that, the secondary component of the direct ejecta from the PNS has more uniform properties, with $Y_e\simeq0.23-0.26$ and $S\simeq10-15~k_{\rm B}$ as well as expansion timescale of $\tau\sim 10$~ms, until $t-t_c\simeq 20$~ms. Consequently, the early component of the direct ejecta produces mainly the first $r$-peak nuclei ($A\sim 90$) through the neutron-rich charged particle freeze-out~\citep{Meyer:1998}, while the second component with lower $Y_e$ can produce nuclei beyond the second peak of $A\sim 130$ and even up to the thrid peak of $A\sim 195$ by the $r$-process.

In contrast, the NDW (blue segments in Fig.~\ref{fig:nucleo_conditions}) features generally high and even increasing entropy $100\lesssim S\lesssim 300$, relatively high and as well increasing $Y_e$ that transits from initially $Y_e\sim 0.45$ early on to higher values of $Y_e\sim 0.5$ at late time.  This evolution of $Y_e$, together with the low and decreasing $\dot M$ from $\lesssim 10^{-2}$~$M_\odot$~s$^{-1}$ to $\sim 10^{-5}$~$M_\odot$~s$^{-1}$, are typical conditions found for the NDW \citep{MartinezPinedo:2012,Roberts:2012,Mirizzi.Tamborra.ea:2016}, except the high entropy.  The high entropy of the wind originates from the high compactness of the PNS, as a direct result from the presence of quark matter at the PNS interior.  Under radiation dominated conditions, the NDW entropy  can be approximated according to Eqs.~(40) and (49) of \citet{Qian:1996xt}, following the Newtonian equations of the steady state NDW expansion.  The expressions feature a direct dependence on the ratio of PNS mass and radius, i.e. the compactness of the nascent PNS.  Such EOS effect on the PNS structure was the cause of the high entropy in the steady-state NDW study of \citet{Wanajo:2001},  see model~C in their Table~I, however, without microscopic model EOS implemented.  There, it was shown that the steady-state NDW, launched from a PNS with mass of $2$~$M_\odot$ and radius of 10~km, can lead to a high entropy of $S\simeq 200~k_{\rm B}$ for neutrino luminosities of $\simeq 10^{51}$~erg~s$^{-1}$, when general relativistic effects are included (see also Fig.~8 of \citet{Otsuki.Tagoshi.ea:2000} and Fig.~11 of \citet{Thompson:2001ys}). This agrees qualitatively with our findings here based on general relativistic radiation-hydrodynamics simulations of the PNS deleptonization. However, we note that our NDW here has in general a higher mass-ejection rate and shorter expansion timescale when compared to those derived in the steady-state wind studies.

In addition to the substantially higher entropy, the slightly neutron-rich conditions with $Y_e\simeq 0.45$ obtained at early times $t-t_c\lesssim 2$~s are related to the slightly larger spectral differences between $\bar\nu_e$ and $\nu_e$ during the early cooling phase (see Fig.~\ref{fig:nuLEa}), compared to our reference hadronic PNS deleptonization (see Fig.~\ref{fig:nuLEb}). Later, material turns less neutron-rich when the spectral differences between all neutrino flavors continues to decrease, as can be identified from Fig.~\ref{fig:nuLEa}. 
As a result, the nucleosynthesis in the NDW proceeds to produce nuclei all the way to the 3$^{\rm rd}$ $r$-process peak and beyond due to the combination of the high entropies $S\simeq 200-300~k_{\rm B}$ and initially slightly neutron-rich conditions ($Y_e\gtrsim 0.45$) for $2\lesssim t-t_c\lesssim 6$~s (see Fig.~\ref{fig:traj}(f).

For the intermediate ejecta (red segments in Fig.~\ref{fig:nucleo_conditions}), they smoothly connect the transition from direct ejecta to the wind and contain
wide ranges of $Y_e$ and $S$ ($0.27\lesssim Y_e\lesssim 0.5$ and $20\lesssim S\lesssim 100$~$k_{\rm B}$).
However, as none of them possesses the right combination of low-enough $Y_e$ or high-enough entropy, 
they only produce nuclei mostly between $50\lesssim A \lesssim 130$ without much beyond the $A\sim 130$ peak. 

\subsection{Dependence on the Progenitor mass}\label{sec:3.3}
We now examine the nucleosynthesis condition and yields from the other two SN explosion models with different progenitor masses of $35$ and 50~$M_\odot$, and compare them with results obtained in the 40~$M_\odot$ case. The orange and magenta thin curves in Fig.~\ref{fig:nucleo_conditions} show the nucleosynthesis relevant quantities in these two models. Both SN explosion models produce qualitatively a similar behavior, as for the 40~$M_\odot$ reference case, in terms of the classification of the ejecta and their corresponding nucleosythesis conditions. Notably, both 35 and 50~$M_\odot$ models produce a slightly lower $Y_e$ in the direct ejecta than that was obtained in the 40~$M_\odot$ model.

The lower $Y_e$ obtained in the 35~$M_\odot$ originates from the weaker explosion which, in turn, results in a lower postshock temperature. Consequently, positron capture operate less effectively to raise $Y_e$. On the other hand, although the $50$~$M_\odot$ model produces the strongest explosion among the three models and its ejecta have the highest postshock temperature, the associated faster expansion compensates this effect such that the obtained $Y_e$ in the direct ejecta is also lower than that in the 40~$M_\odot$ model. For the high-entropy pre-shocked NDW, the $Y_e$ in the 50~$M_\odot$ is visibly smaller than the other two models (see Fig.~\ref{fig:fullst}), as discussed in Sec.~\ref{sec:2.4}. The amount of ejected matter contained in each component for all three models are listed in Table~\ref{tab:ejecta}.

\begin{figure*}[t!]
\centering
\subfigure[Isotopic abundances in comparison to solar (black dots)]{\includegraphics[width=\columnwidth]{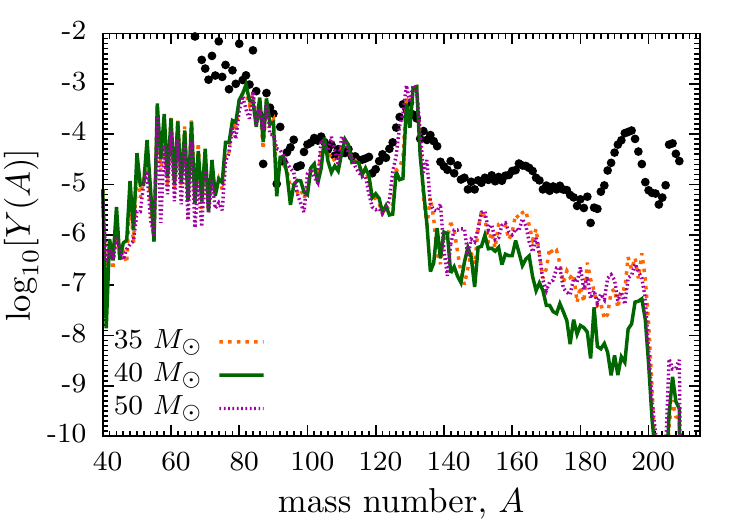}\label{fig:solar}}
\subfigure[Elemental abundances in comparison to metal-poor stars]{\includegraphics[width=\columnwidth]{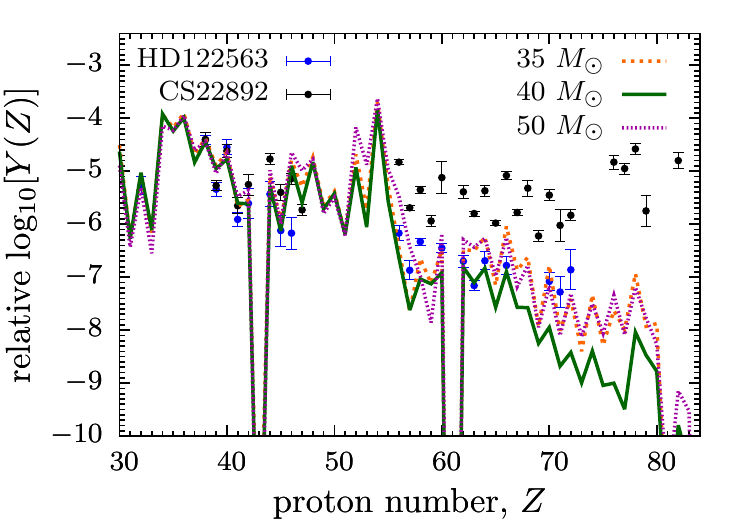}\label{fig:elem}}
\caption{Integrated abundances, $Y$, at the time of 1~Gyr for different SN explosion models listed in Tables~\ref{tab:prog} and \ref{tab:ejecta}. Graph~(a): isotopic abundances in comparison to the solar $r$-process abundances (black dots), scaled to roughly match the abundance peaks at nuclear mass numbers $A\simeq 80$ and $A\simeq 130$. Graph~(b): elemental abundances $Y(Z)$ in comparison to measured abundances of two representative metal-poor stars, HD~122563~\citep{Roederer:2012dr} and CS~22892-052~\citep{Sneden:2003zq}; the abundances are scaled to have the same value at atomic number $Z=38$ (Sr).}
\label{fig:abundances}
\end{figure*}

We show the integrated nucleosynthesis yields of all three ejecta components from all three SN explosion models, compared to the measured Solar $r$ abundance pattern in Fig.~\ref{fig:solar}, scaled to roughly match the abundances at both the first and the second peaks. It clearly shows that this source cannot be the dominating $r$-process site accounting for the entire solar $r$ abundances as the yields beyond the second peak are relatively reduced.  Note that both the 35 and 50~$M_\odot$ models produce larger amount of material above the second peak than the 40~$M_\odot$ model. The enhanced production in between the peaks, $140\lesssim A\lesssim 180$, are due to the lower $Y_e$ obtained in the direct ejecta discussed (see Sec.~\ref{sec:3.2}). On the other hand, the more pronounced third peak in the 35~$M_\odot$ model is driven similarly by the low-$Y_e$ component of the direct ejecta, while in the 50~$M_\odot$ model, it is due to the lower $Y_e$ of the high-entropy NDW (see Fig.~\ref{fig:nucleo_conditions}).

\section{Implications for galactic chemical evolution and deep-sea measurements of heavy elements}
\label{sec:4}
The nucleosynthesis results presented in Section~\ref{sec3} reveal the production of $r$-process elements in phase-transition powered SN explosions. Here we discuss whether these explosions would be able to contribute to the chemical evolution of $r$-process elements, especially at low metallicities, as well as the implications for $^{244}$Pu measurements in deep-ocean crusts.

\subsection{Galactic Chemical Evolution and Low-metallicity Stars}
\label{sec:4.1}
Since many metal-poor stars in the early Galaxy have been found to exhibit abundance patterns similar to the solar $r$ abundances~\citep{Sneden:2008}, a subset of them features under-produced abundances in the rare-earth region corresponding to elements with atomic numbers between $57\leq Z\leq 71$, above the second peak, in comparison to elemental abundances around $Z\sim 40$. The latter refers to the so-called weak-$r$ or limited-$r$ stars \citep[see e.g.][]{Frebel:2018fcn,Cowan.Sneden.ea:2019}. In Fig.~\ref{fig:elem}, we show the deduced $r$-process elemental abundances from our models in comparison to the measured abundances in two representative metal-poor stars, CS22892~\citep{Sneden:2003zq} and HD122563~\citep{Roederer:2012dr}, scaled to match strontium ($Z=38$). As discussed in Sec.~\ref{sec3}, our models under produce heavy elements above the second peak in comparison to observations of CS22892, which represents metal-poor stars with abundance patterns similar to solar; still, they yield qualitatively similar abundances when compared to those observed in HD122563. This suggests that SN explosions powered by the quark-hadron phase transition can be an important source contributing to the $r$-process enrichment in the early Galaxy and can possibly account for those with abundance patterns similar to that of HD~122563. We provide in Table~\ref{tab:ejecta} the produced amount of iron, strontium, and europium for each of the three SN explosion models, which can be used as input in galactic chemical evolution models. Note that in all three SN explosions, both the strontium and europium are predominantly produced in the direct ejecta due to the higher ejecta masses relative to the masses in the NDW (see Table~\ref{tab:ejecta}). Interestingly, the large amount of strontium produced in all models, relative to iron, give rise to $Y(\rm{Sr})/Y(\rm{Fe})$ being a factor of $\simeq 10$ larger than values inferred in metal-poor stars. The excessive amount of strontium may therefore be used to limit the stellar mass range or the fraction of massive stars that explode by the phase transition. A detailed comparison of our models with the observed abundances of strontium and europium in metal-poor stars will be reported in a forthcoming publication.

\subsection{Deep-sea Sediments Containing Heavy Elements}
\label{sec:4.2}
The detection of $^{244}$Pu isotopes in deep-sea sediments \citep{Wallner:2015} suggests the existence of rare nucleosynthesis sites, capable of producing large amounts of actinides, comparable to the typical yield predicted by binary neutron star merger simulations \citep[see also][]{Hotokezaka:2015zea}. This would also apply for alternative rare sources associated with massive stars, such as the magneto-rotationally driven SN \citep[cf.][]{Winteler:2012, Moesta:2014}, or the collapsar accretion disk wind \citep[][]{Siegel:2018zxq}, which remain possible options. Our results complement these considerations and might be a possible explanation for new, improved sensitivity measurements, which suggest that a more frequent source would be required to account for the observed small amount of $^{244}$Pu being co-observed with $^{60}$Fe~\citep[][]{Wallner:2019}: Only core-collapse supernovae can be responsible for the $^{60}$Fe depositions in deep-sea crust samples.\footnote{ Note, however, that the amount of $^{60}$Fe ejected in core-collapse supernovae is rather uncertain. It is mainly constrained from stellar evolution~\citep[see][]{Jones:2019des}.} Other sources, such as AGB stars, Type Ia (thermonuclear) SNe and kilonovae are most likely ruled out to account for the observed $^{60}$Fe yields~\citep[for details, see][and references therein]{Fields:2019}. Analogously, the NDW associated with an {\em ordinary} PNS composed of only hadronic matter fails to provide suitable condition for a strong $r$  process~\citep[see][]{MartinezPinedo:2014,Fischer:2020}. In particular, the associated nucleosyntheiss fails to account for any actinide production. Studies of another potential $r$-process site that may operate in the helium shell of the stellar envelope, aided by neutrino processes, only work in the low metalicity environment~\citep{Banerjee:2011zm} and can thus not be responsible for the recent deposition in deep-sea sediments.

In the light of this work, the high entropy and slightly neutron-rich conditions obtained in the NDW allow for the production of a small amount of actinides. In particular, all three explosion models explored here produce robustly $^{244}$Pu, though of varying amounts of about $10^{-10}-10^{-9}$~$M_\odot$ (see Table~\ref{tab:ejecta}). The 50~$M_\odot$ model produces the largest amount of $^{244}$Pu due to the lowest $Y_e$ obtained in the NDW, which originates from most compact PNS that gives rise to high $\nu_e$ and $\bar\nu_e$ luminosities and average energies, as discussed in previous sections.

The amounts of $^{60}$Fe in our adopted pre-SN models are listed in Table~\ref{tab:ejecta}, being on the order of few times $10^{-5}$~$M_\odot$ \citep[see also][]{Rauscher:2002}. Here, we neglect additional contributions to $^{60}$Fe during the explosive nucleosynthesis phase due to the supernova shock passage as they are expected to be small for high mass progenitors~\citep{Limongi.Chieffi:2006}. However, notice that \citet{Limongi.Chieffi:2006} predict a much larger yield of $^{60}$Fe. Comparing now the $^{60}$Fe and $^{244}$Pu yields from our models results in an abundance ratio of $Y(^{244}\rm Pu)/Y(^{60}\rm Fe)\simeq 10^{-7}-10^{-4}$, making our considerations an interesting scenario for the explanation of the combined occurrence of $^{60}$Fe and $^{244}$Pu in deep-sea sediments, yet this remains to be shown to be compatible with future samples.

\section{Summary}
\label{sec:summary}
In this article, we have discussed the possibility of reviving the {\em classical} site for the $r$-process nucleosynthesis, massive star explosions. To this end, we have performed general relativistic, neutrino-radiation hydrodynamic simulations of core-collapse SN in spherical symmetry, in which a high-density phase transition facilitates the explosions. The simulations are launched from the 35~$M_\odot$ and 40~M$ _\odot$ progenitors of \citet{Rauscher:2002} and the 50~$M_\odot$ progenitor of \citet{Umeda:2007wk}. The associated explosions are enabled due to the phase transition, from normal nuclear matter to a state of deconfined quark matter. The nascent compact PNS contain about 2~$M_\odot$ (baryon mass) and have radii of about 10~km, with a massive quark matter core.

The key aspect for the nucleosynthesis here, in comparison to {\em canonical} core-collapse supernova explosions, is twofold: there is an early ejecta component, here further divided into direct and intermediate ejecta, featuring a short expansion timescale, moderately neutron-rich conditions with $Y_e\simeq 0.25$ to $Y_e=0.45$ and a relatively low entropy per baryon on the order of a few tens of $k_{\rm B}$. This early ejecta component gives rise to the production of heavy elements associated with the first, second and the rare-earth $r$-process peaks, including strontium and europium. The early ejecta are followed by the NDW, which we found here to be dominated by a high-entropy component, again contrary to {\em canonical} NDW from a purely hadronic PNS that result from neutrino-driven SN explosions, due to the massive and compact remnant.

The early direct and intermediate ejecta discussed in the present article show some similarity to the previous work of \citet{Nishimura:2012}, where SN explosions from \citet{Sagert:2008ka} and \citet{Fischer:2011} were analyzed, that considered the hadron-quark phase transition based on the thermodynamic bag model. However, \citet{Nishimura:2012} obtained somewhat less neutron-rich conditions which is why their nucleosynthesis pattern shows a sharp drop of the abundances beyond mass number $A\sim130$. Moreover, the hadron-quark hybrid EOS employed in \citet{Sagert:2008ka} and \citet{Fischer:2011} are in conflict with a variety of constraints, e.g., their maximum hybrid star mass is significantly below the current constraint of about 2~$M_\odot$, resulting form high-precision measurements of massive pulsars in binary systems~\citep{Antoniadis:2013,Cromartie:2019}. This caveat has been overcome in the recent study of \citet{Fischer:2018} developing a new class of hybrid EOS being in agreement with all present nuclear physics and astrophysics constraints.

Furthermore, the subsequent NDW ejected by neutrino heating, not only contain high entropies per baryon, on the order of $200-300~k_{\rm B}$, but are also slightly neutron-rich with $Y_e\simeq 0.45$ for several seconds. This enables the production of nuclei associated with the 3$^{\rm rd}$ $r$-process peak at mass number $A=195$ and even the actinides, including the short-lived radioactive $^{244}$Pu nucleus. However, the relative yields of this component are strongly suppressed compared to the first and second $r$-process peak nuclei produced in the early ejecta.

Although the total $r$-process yields from this new class of SN explosions discussed here do not resemble the main $r$ abundances pattern found in the solar system and in the majority of metal-poor stars, due to the reduced amount of nuclei above the second $r$-process peak, all models produce abundance patterns similar to a subset of metal-poor stars enriched by a weak/limited $r$-process. This agreement indicates that the scenario discussed here can be an interesting source contributing to the $r$-process enrichment in the early galaxies. However, the large amount of the produced strontium relative to iron may limit the fraction of stars in this ZAMS mass range between 35--50~$M_\odot$ to explode in this way. Further detailed galactic chemical evolution studies addressing this issue will be reported in an upcoming article.

Interestingly, the isotopic $^{244}$Pu enrichment found in deep ocean crust samples, being associated with two past nearby SN identified due to the simultaneous $^{60}$Fe enrichment, agrees broadly with the measured ratio of $^{244}$Pu and $^{60}$Fe yields from our predictions. This points to the fact that these events could belong to the present class of rare SN. As a consequence this new site for the $^{244}$Pu measurement is currently under investigation.

It is presently still unclear if quark-matter degrees of freedom appear at high baryon densities that are possibly encountered in the interior of massive neutron stars as well as PNS. Moreover, the nature of such transition at high baryon density is uncertain, e.g., it could be of first-order or a smooth cross over, as predicted by lattice QCD at high temperatures and vanishing baryon density \citep[cf.][]{Laermann:2012PRL,Katz:2014PhLB}. Hence, complementary to the heavy-ion collider programs at FAIR at the GSI in Darmstadt (Germany) and NICA in Dubna (Russia), astrophysics of compact stellar objects has long been explored as a laboratory to probe the hot and dense phases of matter. The explosions of massive supergiant stars due to a first-order phase transition at supranuclear density discussed in this work not only leave an identifiable signal via the emission of a millisecond neutrino burst \citep[][]{Fischer:2018}, but also a unique nucleosynthesis pattern. This opens a new possibility to provide a smoking-gun signature for (or rule out) SN explosions triggered by a (sufficiently strong) hadron-quark phase transition.

\section*{Acknowledgements}
T.F. and N.U.B. acknowledge support from the Polish National Science Center (NCN) under grant Nos. 2016/23/B/ST2/00720 (T.F.), 2019/33/B/ST9/03059 (T.F.) and 2019/32/C/ST2/00556 (N.U.F.B.). M.R.W. acknowledges support from the Academia Sinica by grant No. AS-CDA-109-M11, Ministry of Science and Technology, Taiwan under grant No. 108-2112-M-001-010, and the Physics Division, National Center of Theoretical Science of Taiwan. B.W.  acknowledges the support from the ERC Consolidator Grant (Hungary) funding scheme  (project  RADIOSTAR,  G.A.  n.  724560) and of  the  National  Science  Foundation  (USA)  under  grant  No.  PHY-1430152  (JINA  Center  for  the Evolution  of  the  Elements). G.M.P is partly supported the Deutsche Forschungsgemeinschaft (DFG, German Research Foundation) -- Project-ID 279384907 -- SFB 1245. The supernova simulations were performed at the Wroclaw Center for Scientific Computing and Networking (WCSS) in Wroclaw (Poland). This work was supported by the COST Actions CA16117 ``ChETEC'' and CA16214 ``PHAROS''.

\section*{ORCID iDs}
\small
\noindent Tobias~Fischer~{\color{blue}https://orcid.org/0000-0003-2479-344X}\\
Meng-Ru~Wu~{\color{blue}https://orcid.org/0000-0003-4960-8706}\\
Gabriel~Mart{\'i}nez-Pinedo~{\color{blue} https://orcid.org/0000-0002-
3825-0131} \\
Friedrich-Karl~Thielemann~{\color{blue} https://orcid.org/0000-0002-
7256-9330}


\begin{thebibliography}{0}
\expandafter\ifx\csname natexlab\endcsname\relax\def\natexlab#1{#1}\fi

\end{thebibliography}


\begin{thebibliography}{89}
\expandafter\ifx\csname natexlab\endcsname\relax\def\natexlab#1{#1}\fi

\bibitem[{{Antoniadis} {et~al.}(2013){Antoniadis}, {Freire}, {Wex}, {Tauris},
  {Lynch}, {van Kerkwijk}, {Kramer}, {Bassa}, {Dhillon}, {Driebe}, {Hessels},
  {Kaspi}, {Kondratiev}, {Langer}, {Marsh}, {McLaughlin}, {Pennucci}, {Ransom},
  {Stairs}, {van Leeuwen}, {Verbiest}, \& {Whelan}}]{Antoniadis:2013}
{Antoniadis}, J., {Freire}, P.~C.~C., {Wex}, N., {et~al.} 2013, Science, 340,
  448

\bibitem[{Arcones {et~al.}(2007)Arcones, Janka, \& Scheck}]{Arcones:2006uq}
Arcones, A., Janka, H.-T., \& Scheck, L. 2007, \aap, 467, 1227

\bibitem[{{Arcones} \& {Mart{\'{\i}}nez-Pinedo}(2011)}]{Arcones:2011c}
{Arcones}, A., \& {Mart{\'{\i}}nez-Pinedo}, G. 2011, \prc, 83, 045809

\bibitem[{{Argast} {et~al.}(2004){Argast}, {Samland}, {Thielemann}, \&
  {Qian}}]{Argast:2004}
{Argast}, D., {Samland}, M., {Thielemann}, F.-K., \& {Qian}, Y.-Z. 2004, \aap,
  416, 997

\bibitem[{Banerjee {et~al.}(2011)Banerjee, Haxton, \& Qian}]{Banerjee:2011zm}
Banerjee, P., Haxton, W.~C., \& Qian, Y.-Z. 2011, \prl, 106, 201104

\bibitem[{{Bartl} {et~al.}(2016){Bartl}, {Bollig}, {Janka}, \&
  {Schwenk}}]{Bartl:2016}
{Bartl}, A., {Bollig}, R., {Janka}, H.~T., \& {Schwenk}, A. 2016, \prd, 94,
  083009

\bibitem[{{Bauswein} {et~al.}(2019){Bauswein}, {Bastian}, {Blaschke},
  {Chatziioannou}, {Clark}, {Fischer}, \& {Oertel}}]{Bauswein:2019}
{Bauswein}, A., {Bastian}, N.-U.~F., {Blaschke}, D.~B., {et~al.} 2019, \prl,
  122, 061102

\bibitem[{{Bazavov} {et~al.}(2012){Bazavov}, {Ding}, {Hegde}, {Kaczmarek},
  {Karsch}, {Laermann}, {Mukherjee}, {Petreczky}, {Schmidt}, {Smith},
  {Soeldner}, \& {Wagner}}]{Laermann:2012PRL}
{Bazavov}, A., {Ding}, H.-T., {Hegde}, P., {et~al.} 2012, \prl, 109, 192302

\bibitem[{Benic {et~al.}(2015)Benic, Blaschke, Alvarez-Castillo, Fischer, \&
  Typel}]{Benic:2014jia}
Benic, S., Blaschke, D., Alvarez-Castillo, D.~E., Fischer, T., \& Typel, S.
  2015, \aap, 577, A40

\bibitem[{{Bors{\'a}nyi}{et~al.}(2014)
{Bors{\'a}nyi}, {Fodor}, {Hoelbling}, {Katz}, {Krieg}, \& {Szab{\'o}}}]{Katz:2014PhLB}
{Bors{\'a}nyi}, S., {Fodor}, Z., {Hoelbling}, C., {et~al.} 2014, Phys. Lett. B,
  730, 99

\bibitem[{{Burrows} \& {Sawyer}(1998)}]{Burrows:1998}
{Burrows}, A., \& {Sawyer}, R.~F. 1998, \prc, 58, 554

\bibitem[{C{\^o}t{\'e} {et~al.}(2019)
{{C{\^o}t{\'e}}, Benoit and {Eichler}, Marius and {Arcones}, Almudena and {Hansen}, Camilla J. and {Simonetti}, Paolo and {Frebel}, Anna and      {Fryer}, Chris L. and {Pignatari}, Marco and {Reichert}, Moritz and {Belczynski}, Krzysztof and {Matteucci}, Francesca}}]{Cote:2018qku}
C{\^o}t{\'e}, B., {et~al.} 2019, \apj, 875, 106

\bibitem[{{Cowan} \& {Sneden}(2006)}]{Cowan:2006}
{Cowan}, J.~J., \& {Sneden}, C. 2006, \nat, 440, 1151

\bibitem[{{Cowan} {et~al.}(2019){Cowan}, {Sneden}, {Lawler}, {Aprahamian},
  {Wiescher}, {Langanke}, {Mart{\'\i}nez-Pinedo}, \&
  {Thielemann}}]{Cowan.Sneden.ea:2019}
{Cowan}, J.~J., {Sneden}, C., {Lawler}, J.~E., {et~al.} 2019, arXiv e-prints,
  arXiv:1901.01410

\bibitem[{{Cromartie} {et~al.}(2020){Cromartie}, {Fonseca}, {Ransom},
  {Demorest}, {Arzoumanian}, {Blumer}, {Brook}, {DeCesar}, {Dolch}, {Ellis},
  {Ferdman}, {Ferrara}, {Garver-Daniels}, {Gentile}, {Jones}, {Lam}, {Lorimer},
  {Lynch}, {McLaughlin}, {Ng}, {Nice}, {Pennucci}, {Spiewak}, {Stairs},
  {Stovall}, {Swiggum}, \& {Zhu}}]{Cromartie:2019}
{Cromartie}, H.~T., {Fonseca}, E., {Ransom}, S.~M., {et~al.} 2020, \nat~Astronomy, 4, 72

\bibitem[{Fernández \& Metzger(2013)}]{Fernandez:2013tya}
Fernández, R., \& Metzger, B.~D. 2013, \mnras, 435, 502

\bibitem[{{Fields} {et~al.}(2019){Fields}, {Ellis}, {Binns}, {Breitschwerdt},
  {deNolfo}, {Diehl}, {Dwarkadas}, {Ertel}, {Faestermann}, {Feige}, {Fitoussi},
  {Frisch}, {Graham}, {Haley}, {Heger}, {Hillebrand t}, {Israel}, {Janka},
  {Kachelrei{\ss}}, {Korschinek}, {Limongi}, {Lugaro}, {Marinho}, {Melott},
  {Mewaldt}, {Miller}, {Ogliore}, {Paul}, {Paulucci}, {Pecaut}, {Rauch},
  {Rehm}, {Schulreich}, {Seitenzahl}, {Sorensen}, {Thielemann}, {Timmes},
  {Thomas}, \& {Wallner}}]{Fields:2019}
{Fields}, B., {Ellis}, J.~R., {Binns}, W.~R., {et~al.} 2019, \baas, 51, 410

\bibitem[{Fischer {et~al.}(2010)Fischer, Whitehouse, Mezzacappa, Thielemann, \&
  Liebend{\"o}rfer}]{Fischer:2009af}
Fischer, T., Whitehouse, S., Mezzacappa, A., Thielemann, F.-K., \&
  Liebend{\"o}rfer, M. 2010, \aap, 517, A80

\bibitem[{{Fischer} {et~al.}(2011){Fischer}, {Sagert}, {Pagliara}, {Hempel},
  {Schaffner-Bielich}, {Rauscher}, {Thielemann}, {K{\"a}ppeli},
  {Mart{\'{\i}}nez-Pinedo}, \& {Liebend{\"o}rfer}}]{Fischer:2011}
{Fischer}, T., {Sagert}, I., {Pagliara}, G., {et~al.} 2011, \apjs, 194, 39

\bibitem[{{Fischer}(2016)}]{Fischer:2016b}
{Fischer}, T. 2016, \aap, 593, A103

\bibitem[{{Fischer} {et~al.}(2018){Fischer}, {Bastian}, {Wu}, {Baklanov},
  {Sorokina}, {Blinnikov}, {Typel}, {Kl{\"a}hn}, \& {Blaschke}}]{Fischer:2018}
{Fischer}, T., {Bastian}, N.-U.~F., {Wu}, M.-R., {et~al.} 2018, \nat~Astron., 2, 980

\bibitem[{{Fischer} {et~al.}(2020){Fischer}, {Guo}, {Dzhioev},
  {Mart{\'\i}nez-Pinedo}, {Wu}, {Lohs}, \& {Qian}}]{Fischer:2020}
{Fischer}, T., {Guo}, G., {Dzhioev}, A.~A., {et~al.} 2020, \prc, 101, 025804

\bibitem[{{Fonseca} {et~al.}(2016){Fonseca}, {Pennucci}, {Ellis}, {Stairs},
  {Nice}, {Ransom}, {Demorest}, {Arzoumanian}, {Crowter}, {Dolch}, {Ferdman},
  {Gonzalez}, {Jones}, {Jones}, {Lam}, {Levin}, {McLaughlin}, {Stovall},
  {Swiggum}, \& {Zhu}}]{Fonseca:2016}
{Fonseca}, E., {Pennucci}, T.~T., {Ellis}, J.~A., {et~al.} 2016, \apj, 832, 167

\bibitem[{Frebel(2018)}]{Frebel:2018fcn}
Frebel, A. 2018, Ann. Rev. Nucl. Part. Sci., 68, 237

\bibitem[{Fr{\"o}hlich {et~al.}(2006)Fr{\"o}hlich, Martinez-Pinedo,
  Liebendorfer, Thielemann, Bravo, {et~al.}}]{Froehlich:2005ys}
Fr{\"o}hlich, C., Martinez-Pinedo, G., Liebend{\"o}rfer, M., {et~al.} 2006, \prl,
  96, 142502

\bibitem[{{Goriely} {et~al.}(2011){Goriely}, {Bauswein}, \&
  {Janka}}]{Goriely:2011}
{Goriely}, S., {Bauswein}, A., \& {Janka}, H.-T. 2011, \apjl, 738, L32

\bibitem[{Hempel {et~al.}(2009)Hempel, Pagliara, \&
  Schaffner-Bielich}]{Hempel:2009vp}
Hempel, M., Pagliara, G., \& Schaffner-Bielich, J. 2009, \prd, 80, 125014

\bibitem[{{Hirai} {et~al.}(2015){Hirai}, {Ishimaru}, {Saitoh}, {Fujii},
  {Hidaka}, \& {Kajino}}]{Hirai:2015}
{Hirai}, Y., {Ishimaru}, Y., {Saitoh}, T.~R., {et~al.} 2015, \apj, 814, 41

\bibitem[{Hoffman {et~al.}(1997)Hoffman, Woosley, \& Qian}]{Hoffman:1996aj}
Hoffman, R., Woosley, S., \& Qian, Y. 1997, \apj, 482, 951

\bibitem[{Horowitz(2002)}]{Horowitz:2001xf}
Horowitz, C. 2002, \prd, 65, 043001

\bibitem[{{Horowitz} {et~al.}(1985){Horowitz}, {Moniz}, \&
  {Negele}}]{Horowitz:1985}
{Horowitz}, C.~J., {Moniz}, E.~J., \& {Negele}, J.~W. 1985, \prd, 31, 1689

\bibitem[{Hotokezaka {et~al.}(2018)Hotokezaka, Beniamini, \&
  Piran}]{Hotokezaka:2018aui}
Hotokezaka, K., Beniamini, P., \& Piran, T. 2018, Int. J. Mod. Phys., D27,
  1842005

\bibitem[{Hotokezaka {et~al.}(2015)Hotokezaka, Piran, \&
  Paul}]{Hotokezaka:2015zea}
Hotokezaka, K., Piran, T., \& Paul, M. 2015, Nature Phys., 11, 1042

\bibitem[{H{\"u}depohl {et~al.}(2010)H{\"u}depohl, M{\"u}ller, Janka, Marek, \&
  Raffelt}]{Huedepohl:2010}
H{\"u}depohl, L., M{\"u}ller, B., Janka, H.-T., Marek, A., \& Raffelt, G.~G.
  2010, \prl, 104, 251101

\bibitem[{Jones {et~al.}(2019)Jones, Möller, Fryer, Fontes, Trappitsch, Even,
  Couture, Mumpower, \& Safi-Harb}]{Jones:2019des}
Jones, S.~W., Möller, H., Fryer, C.~L., {et~al.} 2019, \mnras, 485, 4287

\bibitem[{Just {et~al.}(2015)Just, Bauswein, Pulpillo, Goriely, \&
  Janka}]{Just:2014fka}
Just, O., Bauswein, A., Pulpillo, R.~A., Goriely, S., \& Janka, H.~T. 2015,
  \mnras, 448, 541

\bibitem[{{Kaltenborn} {et~al.}(2017){Kaltenborn}, {Bastian}, \&
  {Blaschke}}]{Blaschke:2017}
{Kaltenborn}, M.~A.~R., {Bastian}, N.-U.~F., \& {Blaschke}, D.~B. 2017, \prd,
  96, 056024

\bibitem[{Kitaura {et~al.}(2006)Kitaura, Janka, \& Hillebrandt}]{Kitaura:2006}
Kitaura, F., Janka, H.-T., \& Hillebrandt, W. 2006, \aap, 450, 345

\bibitem[{{Kl{\"a}hn} \& {Fischer}(2015)}]{Klaehn:2015}
{Kl{\"a}hn}, T., \& {Fischer}, T. 2015, \apj, 810, 134

\bibitem[{{Komiya} \& {Shigeyama}(2016)}]{Komiya:2016xx}
{Komiya}, Y., \& {Shigeyama}, T. 2016, \apj, 830, 76

\bibitem[{Korobkin {et~al.}(2012)Korobkin, Rosswog, Arcones, \&
  Winteler}]{Korobkin:2012uy}
Korobkin, O., Rosswog, S., Arcones, A., \& Winteler, C. 2012, \mnras, 426, 1940

\bibitem[{Liebend\"orfer {et~al.}(2004)Liebend\"orfer, Messer, Mezzacappa,
  Bruenn, Cardall, {et~al.}}]{Liebendoerfer:2004}
Liebend\"orfer, M., Messer, O., Mezzacappa, A., {et~al.} 2004, \apjs, 150, 263

\bibitem[{Limongi \& Chieffi(2006)}]{Limongi.Chieffi:2006}
Limongi, M., \& Chieffi, A. 2006, \apj, 647, 483

\bibitem[{{Mart{\'{\i}}nez-Pinedo} {et~al.}(2012){Mart{\'{\i}}nez-Pinedo},
  {Fischer}, {Lohs}, \& {Huther}}]{MartinezPinedo:2012}
{Mart{\'{\i}}nez-Pinedo}, G., {Fischer}, T., {Lohs}, A., \& {Huther}, L. 2012,
  \prl, 109, 251104

\bibitem[{{Mart{\'{\i}}nez-Pinedo} {et~al.}(2014){Mart{\'{\i}}nez-Pinedo},
  {Fischer}, \& {Huther}}]{MartinezPinedo:2014}
{Mart{\'{\i}}nez-Pinedo}, G., {Fischer}, T., \& {Huther}, L. 2014, Journal of
  Physics G -- Nuclear Physics, 41, 044008

\bibitem[{{Maslov} {et~al.}(2019){Maslov}, {Yasutake}, {Ayrian}, {Blaschke},
  {Maruyama}, {Tatsumi}, \& {Voskresensky}}]{Yasutake:2019}
{Maslov}, K., {Yasutake}, N., {Ayrian}, A., {et~al.} 2019, \prc, 100, 025802

\bibitem[{Mendoza-Temis {et~al.}(2015)Mendoza-Temis, Wu, Martínez-Pinedo,
  Langanke, Bauswein, \& Janka}]{Mendoza-Temis:2014mja}
Mendoza-Temis, J., Wu, M.-R., Martínez-Pinedo, G., {et~al.} 2015, \prc,
  92, 055805

\bibitem[{{Meyer} {et~al.}(1998){Meyer}, {Krishnan}, \& {Clayton}}]{Meyer:1998}
{Meyer}, B.~S., {Krishnan}, T.~D., \& {Clayton}, D.~D. 1998, \apj, 498, 808

\bibitem[{Mezzacappa \& Bruenn(1993)}]{Mezzacappa:1993gm}
Mezzacappa, A., \& Bruenn, S. 1993, \apj, 405, 637

\bibitem[{{Mirizzi} {et~al.}(2016){Mirizzi}, {Tamborra}, {Janka}, {Saviano},
  {Scholberg}, {Bollig}, {H{\"u}depohl}, \&
  {Chakraborty}}]{Mirizzi.Tamborra.ea:2016}
{Mirizzi}, A., {Tamborra}, I., {Janka}, H.-T., {et~al.} 2016, Riv. del Nuovo
  Cim., 39, 1

\bibitem[{Moller {et~al.}(1995)Moller, Nix, Myers, \&
  Swiatecki}]{Moller:1993ed}
Moller, P., Nix, J.~R., Myers, W.~D., \& Swiatecki, W.~J. 1995, Atom. Data
  Nucl. Data Tabl., 59, 185

\bibitem[{Moller {et~al.}(2003)Moller, Pfeiffer, \& Kratz}]{Moller:2003fn}
Moller, P., Pfeiffer, B., \& Kratz, K.-L. 2003, \prc, 67, 055802

\bibitem[{{M{\"o}sta} {et~al.}(2014){M{\"o}sta}, {Richers}, {Ott}, {Haas},
  {Piro}, {Boydstun}, {Abdikamalov}, {Reisswig}, \& {Schnetter}}]{Moesta:2014}
{M{\"o}sta}, P., {Richers}, S., {Ott}, C.~D., {et~al.} 2014, \apjl, 785, L29

\bibitem[{{Nishimura} {et~al.}(2012){Nishimura}, {Fischer}, {Thielemann},
  {Fr{\"o}hlich}, {Hempel}, {K{\"a}ppeli}, {Mart{\'i}nez-Pinedo}, \&
  {Rauscher}}]{Nishimura:2012}
{Nishimura}, N., {Fischer}, T., {Thielemann}, F.-K., {et~al.} 2012, \apj, 758, 13

\bibitem[{Otsuki {et~al.}(2000)Otsuki, Tagoshi, Kajino, \&
  Wanajo}]{Otsuki.Tagoshi.ea:2000}
Otsuki, K., Tagoshi, H., Kajino, T., \& Wanajo, S. 2000, \apj, 533, 424

\bibitem[{{Pruet} {et~al.}(2006){Pruet}, {Hoffman}, {Woosley}, {Janka}, \&
  {Buras}}]{Pruet:2006}
{Pruet}, J., {Hoffman}, R.~D., {Woosley}, S.~E., {Janka}, H.-T., \& {Buras}, R.
  2006, \apj, 644, 1028

\bibitem[{{Pruet} {et~al.}(2005){Pruet}, {Woosley}, {Buras}, {Janka}, \&
  {Hoffman}}]{Pruet:2005}
{Pruet}, J., {Woosley}, S.~E., {Buras}, R., {Janka}, H.-T., \& {Hoffman}, R.~D.
  2005, \apj, 623, 325

\bibitem[{Qian \& Woosley(1996)}]{Qian:1996xt}
Qian, Y., \& Woosley, S. 1996, \apj, 471, 331

\bibitem[{{Qian} \& {Wasserburg}(2007)}]{Qian.Wasserburg:2007}
{Qian}, Y.-Z., \& {Wasserburg}, G.~J. 2007, \prep, 442, 237

\bibitem[{{Radice} {et~al.}(2016){Radice}, {Galeazzi}, {Lippuner}, {Roberts},
  {Ott}, \& {Rezzolla}}]{Radice:2016}
{Radice}, D., {Galeazzi}, F., {Lippuner}, J., {et~al.} 2016, \mnras, 460, 3255

\bibitem[{{Rauscher} {et~al.}(2002){Rauscher}, {Heger}, {Hoffman}, \&
  {Woosley}}]{Rauscher:2002}
{Rauscher}, T., {Heger}, A., {Hoffman}, R.~D., \& {Woosley}, S.~E. 2002, \apj,
  576, 323

\bibitem[{Reddy {et~al.}(1998)Reddy, Prakash, \& Lattimer}]{Reddy:1998}
Reddy, S., Prakash, M., \& Lattimer, J.~M. 1998, \prd, 58, 013009

\bibitem[{{Roberts} {et~al.}(2012){Roberts}, {Reddy}, \& {Shen}}]{Roberts:2012}
{Roberts}, L.~F., {Reddy}, S., \& {Shen}, G. 2012, \prc, 86, 065803

\bibitem[{Roederer {et~al.}(2012)Roederer, Lawler, Sobeck, Beers, Cowan,
  Frebel, Ivans, Schatz, Sneden, \& Thompson}]{Roederer:2012dr}
Roederer, I.~U., Lawler, J.~E., Sobeck, J.~S., {et~al.} 2012, \apjs, 203, 27

\bibitem[{{R{\"o}pke} {et~al.}(1986){R{\"o}pke}, {Blaschke}, \&
  {Schulz}}]{Blaschke:1986}
{R{\"o}pke}, G., {Blaschke}, D., \& {Schulz}, H. 1986, \prd, 34, 3499

\bibitem[{Sagert {et~al.}(2009)Sagert, Fischer, Hempel, Pagliara,
  Schaffner-Bielich, {et~al.}}]{Sagert:2008ka}
Sagert, I., Fischer, T., Hempel, M., {et~al.} 2009, \prl, 102, 081101

\bibitem[{Siegel \& Metzger(2017)}]{Siegel:2017nub}
Siegel, D.~M., \& Metzger, B.~D. 2017, \prl, 119, 231102

\bibitem[{Siegel {et~al.}(2018)Siegel, Barnes, \& Metzger}]{Siegel:2018zxq}
Siegel, D.~M., Barnes, J., \& Metzger, B.~D. 2019, \nat, 569, 241

\bibitem[{{Sneden} {et~al.}(2008){Sneden}, {Cowan}, \& {Gallino}}]{Sneden:2008}
{Sneden}, C., {Cowan}, J.~J., \& {Gallino}, R. 2008, \araa, 46, 241

\bibitem[{Sneden {et~al.}(2003){Sneden}, {Cowan}, {Lawler}, {Ivans}, {Burles}, {Beers}, {Primas}, {Hill}, {Truran}, {Fuller}, {Pfeiffer}, {Kratz}, }]{Sneden:2003zq}
Sneden, C., {et~al.} 2003, \apj, 591, 936

\bibitem[{Takahashi {et~al.}(1994)Takahashi, Witti, \&
  Janka}]{Takahashi:1994yz}
Takahashi, K., Witti, J., \& Janka, H.-T. 1994, \aap, 286, 857

\bibitem[{Thompson {et~al.}(2001)Thompson, Burrows, \& Meyer}]{Thompson:2001ys}
Thompson, T.~A., Burrows, A., \& Meyer, B.~S. 2001, \apj, 562, 887

\bibitem[{Typel {et~al.}(2010)Typel, Ropke, Klahn, Blaschke, \&
  Wolter}]{Typel:2009sy}
Typel, S., R{\"o}pke, G., Kl{\"a}hn, T., Blaschke, D., \& Wolter, H. 2010, \prc, 81, 015803

\bibitem[{Umeda \& Nomoto(2008)}]{Umeda:2007wk}
Umeda, H., \& Nomoto, K. 2008, \apj, 673, 1014

\bibitem[{{Wallner}(2019)}]{Wallner:2019}
{Wallner}, A. 2019, private communication

\bibitem[{{Wallner} {et~al.}(2015){Wallner}, {Faestermann}, {Feige},
  {Feldstein}, {Knie}, {Korschinek}, {Kutschera}, {Ofan}, {Paul}, {Quinto},
  {Rugel}, \& {Steier}}]{Wallner:2015}
{Wallner}, A., {Faestermann}, T., {Feige}, J., {et~al.} 2015, \nat~Comm., 6, 5956

\bibitem[{{Wanajo} {et~al.}(2011){Wanajo}, {Janka}, \&
  {M{\"u}ller}}]{Wanajo:2011a}
{Wanajo}, S., {Janka}, H.-T., \& {M{\"u}ller}, B. 2011, \apjl, 726, L15+

\bibitem[{{Wanajo} {et~al.}(2001){Wanajo}, {Kajino}, {Mathews}, \&
  {Otsuki}}]{Wanajo:2001}
{Wanajo}, S., {Kajino}, T., {Mathews}, G.~J., \& {Otsuki}, K. 2001, \apj, 554,
  578

\bibitem[{{Wanajo} {et~al.}(2009){Wanajo}, {Nomoto}, {Janka}, {Kitaura}, \&
  {M{\"u}ller}}]{Wanajo:2009}
{Wanajo}, S., {Nomoto}, K., {Janka}, H.-T., {Kitaura}, F.~S., \& {M{\"u}ller},
  B. 2009, \apj, 695, 208

\bibitem[{Wanajo {et~al.}(2014)Wanajo, Sekiguchi, Nishimura, Kiuchi, Kyutoku,
  \& Shibata}]{Wanajo:2014wha}
Wanajo, S., Sekiguchi, Y., Nishimura, N., {et~al.} 2014, \apj, 789, L39

\bibitem[{{Wehmeyer} {et~al.}(2015){Wehmeyer}, {Pignatari}, \&
  {Thielemann}}]{Wehmeyer:2015}
{Wehmeyer}, B., {Pignatari}, M., \& {Thielemann}, F.-K. 2015, \mnras, 452, 1970

\bibitem[{{Winteler} {et~al.}(2012){Winteler}, {K{\"a}ppeli}, {Perego},
  {Arcones}, {Vasset}, {Nishimura}, {Liebend{\"o}rfer}, \&
  {Thielemann}}]{Winteler:2012}
{Winteler}, C., {K{\"a}ppeli}, R., {Perego}, A., {et~al.} 2012, \apjl, 750, L22

\bibitem[{{Witti} {et~al.}(1994){Witti}, {Janka}, \& {Takahashi}}]{Witti:1994}
{Witti}, J., {Janka}, H.-T., \& {Takahashi}, K. 1994, \aap, 286, 841

\bibitem[{Woosley {et~al.}(2002)Woosley, Heger, \& Weaver}]{Woosley:2002zz}
Woosley, S., Heger, A., \& Weaver, T. 2002, Rev.Mod.Phys., 74, 1015

\bibitem[{Woosley \& Weaver(1995)}]{Woosley:1995ip}
Woosley, S., \& Weaver, T. 1995, \apjs, 101, 181

\bibitem[{Woosley {et~al.}(1994)Woosley, Wilson, Mathews, Hoffman, \&
  Meyer}]{Woosley:1994ux}
Woosley, S., Wilson, J., Mathews, G., Hoffman, R., \& Meyer, B. 1994, \apj,
  433, 229

\bibitem[{Wu {et~al.}(2019)Wu, Barnes, Martinez-Pinedo, \&
  Metzger}]{Wu:2018mvg}
Wu, M.-R., Barnes, J., Martinez-Pinedo, G., \& Metzger, B.~D. 2019, \prl, 122,
  062701

\bibitem[{Wu {et~al.}(2016)Wu, Fernández, Martínez-Pinedo, \&
  Metzger}]{Wu:2016pnw}
Wu, M.-R., Fernández, R., Martínez-Pinedo, G., \& Metzger, B.~D. 2016, \mnras, 463, 2323

\bibitem[{{Yasutake} {et~al.}(2014){Yasutake}, {{\L}astowiecki}, {Beni{\'c}},
  {Blaschke}, {Maruyama}, \& {Tatsumi}}]{Yasutake:2014}
{Yasutake}, N., {{\L}astowiecki}, R., {Beni{\'c}}, S., {et~al.} 2014, \prc, 89,
  065803

\end{thebibliography}
\end{document}